\documentclass[twocolumn]{pasj01}

\draft
\usepackage{color}
\usepackage{booktabs} 
\usepackage{bm} 

\newcommand{\icarus}{Icarus}

\begin{document}

\title{Accelerated FDPS --- Algorithms to Use Accelerators with FDPS}

\author{Masaki \textsc{Iwasawa}\altaffilmark{1}}
\email{masaki.iwasawa@riken.jp}

\author{Daisuke \textsc{Namekata}\altaffilmark{1}}
\email{daisuke.namekata@riken.jp}

\author{Keigo \textsc{Nitadori}\altaffilmark{1}}
\email{keigo@riken.jp}

\author{Kentaro \textsc{Nomura}\altaffilmark{1}}
\email{kentaro.nomura@riken.jp}

\author{Long \textsc{Wang}\altaffilmark{2,1}}
\email{long.wang@riken.jp}

\author{Miyuki \textsc{Tsubouchi}\altaffilmark{1}}
\email{miyuki.tsubouchi@riken.jp}

\author{Junichiro \textsc{Makino}\altaffilmark{3,1,4}}
\email{makino@mail.jmlab.jp}

\altaffiltext{1}{RIKEN Center for Computational Science}

\altaffiltext{2}{Helmholtz Institut f\"{u}r Strahlen und Kernphysik}

\altaffiltext{3}{Department of Planetology, Graduate School of
  Science, Kobe University}

\altaffiltext{4}{Earth-Life Science Institute, Tokyo Institute of
  Technology}


\maketitle

\begin{abstract}

  In this paper, we describe the algorithms we implemented in FDPS
  (Framework for Developing Particle Simulators) to make efficient use
  of accelerator hardware such as GPGPUs (General-purpose computing on
  graphics processing units). We have developed FDPS to make it
  possible for many researchers to develop their own high-performance
  parallel particle-based simulation programs without spending large
  amount of time for parallelization and performance tuning.  The
  basic idea of FDPS is to provide a high-performance implementation
  of parallel algorithms for particle-based simulations in a
  ``generic'' form, so that researchers can define their own particle
  data structure and interparticle interaction functions and supply
  them to FDPS. FDPS compiled with user-supplied data type and
  interaction function provides all necessary functions for
  parallelization, and using those functions researchers can write
  their programs as though they are writing simple non-parallel
  program. It has been possible to use accelerators with FDPS, by
  writing the interaction function that uses the accelerator. However,
  the efficiency was limited by the latency and bandwidth of
  communication between the CPU and the accelerator and also by the
  mismatch between the available degree of parallelism of the
  interaction function and that of the hardware parallelism. We have
  modified the interface of user-provided interaction function so that
  accelerators are more efficiently used. We also implemented new
  techniques which reduce the amount of work on the side of CPU and
  amount of communication between CPU and accelerators. We have
  measured the performance of N-body simulations on a systems with
  NVIDIA Volta GPGPU using FDPS and the achieved performance is around
  27 \% of the theoretical peak limit. We have constructed a detailed
  performance model, and found that the current implementation can
  achieve good performance on systems with much smaller memory and
  communication bandwidth. Thus our implementation will be good for
  future generations of accelerator system.
  
\end{abstract}

\section{Introduction}
\label{sect:intro}

In this paper, we describe new algorithms we implemented in FDPS
[Framework for Developing Particle Simulators, \citep{Iwasawaetal2016,
    2018PASJ..tmp...76N}], to make efficient use of accelerators such
as GPGPUs (General-purpose computing on graphics processing
units). FDPS is designed to make it easy for many researchers to
develop their own programs for particle-based simulations. To develop
efficient parallel programs for particle-based simulations requires a
very large amount of work, like the work of a large team of people for
many years. This is of course true not only for particle-based
simulations, but practically for any large-scale parallel applications
in computational science. The main cause for this problem is that
modern HPC (high-performance computing) platforms have become very
complex, and thus it requires lots of efforts to develop complex
programs to make efficient use of such platforms.

Typical modern HPC systems are actually a cluster of computing nodes
connected through a network, each with typically one or two processor
chips. Largest systems at present consists of around $10^5$ nodes, and
we will see even larger systems soon. This extremely large number of
nodes has made the design of inter-node network very difficult, and
the design of parallel algorithm also has become very difficult. The
calculation times of all nodes must be accurately balanced. The time
necessary for communication must be small enough so that the use of
large systems is meaningful. The communication bandwidth between nodes
is much lower than the main memory bandwidth, which itself is very
small compared to the calculation speed of CPUs. Thus, it is crucial
to avoid communications as much as possible. The calculation time can
show increase, instead of showing decrease as it should, when we use a
large number of nodes, unless we are careful to achieve good load
balance between nodes and to minimize communication.

In addition, the programming environments available on present-day
parallel systems are not easy to use. What is most widely used is MPI,
with which we need to write explicitly how each node communicates with
others in the system. Just to write and debug the program is
difficult, and it has become nearly impossible for any single person
or even for a small group of people to develop large-scale simulation
programs which run efficiently on modern HPC systems.

Moreover, this extremely large number of nodes is just one of the many
difficulties of using modern HPC systems, since within one node, there
are many levels of parallelisms which should be taken care of by
programmers. To make the matter even more complicated, these multiple
levels of parallelism are interwoven with multiple levels of memory
hierarchy with varying bandwidth and latency. For example, the
supercomputer Fugaku, which is under development in Japan as of the
time of writing, will have 48 CPUs (cores) in one chip. These 48 cores
are divided into four groups, each with 12 cores. Cores in one group
share one level-2 cache memory. The cache memories in different groups
communicate with each other through cache-coherency protocol. Thus,
the access of one core to the data which happens to be in its level-2
cache is fast, but that in the cache of another group can be very
slow. Also, the access to the main memory is much slower, and that to
local level-1 cache is much faster. Thus, we need to take into account
the number of cores and the size and speed of caches at each level to
achieve an acceptable performance. To make the matter even worse, many
of modern microprocessors have level-3 or even level-4 caches.

As the result of these difficulties, only a small number of
researchers (or groups of researchers) can develop their own
simulation programs. In the case of cosmological and galactic $N$-body
and SPH (Smoothed Particle Hydrodynamics) simulations, Gadget
\citep{2005MNRAS.364.1105S} and {\tt pkdgrav} \citep{Stadel2001} are
most widely used. For star cluster simulations, NBODY6++ (and
NBODY6++GPU) \citep{2012MNRAS.424..545N} is effectively the
standard. For planetary ring dynamics, REBOUND \citep{ReinLiu2012} has
been available. There has been no public code for simulations of
planetary formation process until recently.

This situation is clearly unhealthy. In many cases, the physics needs
to be modeled is quite simple: particles interact through gravity, and
with some other interactions such as physical collisions. Even so,
almost all researchers are now forced to use existing programs
developed by someone else, simply because HPC platforms have become
too difficult to use. To add some functionality which is not already
implemented in existing programs can be very difficult. In order to
make it possible for researchers to develop their own parallel codes
for particle-based simulations, we have developed FDPS
\citep{Iwasawaetal2016}.

The basic idea of FDPS is to separate the code for parallelization and
that for interaction calculation and numerical integration. FDPS
provides the library functions necessary for parallelization, and
using them researchers write programs very similar to what they would
write for single CPU. Parallelization on multiple nodes and on
multiple cores in single node are taken care of by FDPS.

FDPS provides three sets of functions. One is for the domain
decomposition. Given the data of particles in each nodes, FDPS
performs the decomposition of the computational domain. The decomposed
domains are assigned to MPI processes. The second one is to let MPI
processes exchange particles. Each particle should be sent to the
appropriate MPI process. The third set of functions perform the
interaction calculation. FDPS uses parallel version of Barnes-Hut
algorithm, for both of long-range interactions such as gravitational
interactions and short-range interactions such as inter-molecular
force or fluid interaction. Application program gives the function to
perform interaction calculation for two groups of particles (one group
exerting forces to the other), and FDPS calculates the interaction
using that function.

FDPS offers very good performance on large-scale parallel systems
consisting of ``homogeneous'' multi-core processors, such as K
computer and Cray systems based on x86 processors. On the other hand,
the architecture of large-scale HPC systems is moving from homogeneous
multi-core processors to accelerator-based systems and heterogeneous
multi-core processors.

GPGPUs are most widely used accelerators, and are available on many
large-scale systems. They offer the price-performance ratios and
performance per watt numbers significantly better than those of
homogeneous systems, primarily by integrating a large number of
relatively simple processors on single accelerator chip. On the other
hand, accelerator-based systems have two problems. One is that for
many applications, the communication bandwidth between CPUs and
accelerators becomes the bottleneck. The second one is that because
CPUs and accelerators have separate memory spaces, the programming is
complicated and we cannot use existing programs.

Though in general it is difficult to use accelerators, for
particle-based simulations the efficient use of accelerators is not so
difficult, and that fact is the reason why GRAPE families of
accelerators specialized for gravitational $N$-body simulations had
been successful \citep{2003PASJ...55.1163M}.  GPGPUs are also widely
used both for collisional \citep{2009NewA...14..630G} and
collisionless \citep{2012JCoPh.231.2825B} gravitational $N$-body
simulations. Thus, it is clearly desirable for FDPS to support
accelerator-based architectures.

Though gravitational $N$-body simulation codes have achieved very good
performance on large clusters of GPGPUs, to achieve high efficiency
for particle systems with short-range interactions is difficult. For
example, there exist many high-performance implementations of SPH
algorithms on single GPGPU, or relatively small number of multiple
GPGPUs (around six), but there are not many high-performance SPH codes
for large-scale parallel GPGPU systems. Practically all efficient
GPGPU implementations of SPH algorithm uses GPGPU to run the entire
simulation code, in order to eliminate the communication overhead of
GPGPUs and CPUs. The calculation cost of particle-particle
interactions dominates the total calculation cost of SPH
simulations. Thus, as far as the calculation cost is concerned, it is
sufficient to let GPGPUs evaluate the interactions, and let CPUs
perform the rest of the calculation. However, because of relatively
low communication bandwidth between CPUs and GPGPUs, we need to avoid
the data transfer between them, and if we let GPGPUs do all
calculations, it is clear that we can minimize the communication.

On the other hand, it is more difficult to develop programs for GPGPUs
than for CPUs, and to develop MPI parallel programs for multiple GPGPUs
is clearly more difficult. To make such MPI parallel program for
GPGPUs run on a large cluster is close to impossible.

In order to add the support of GPGPU and other accelerators to FDPS,
we decided to take a different approach. We keep the simple model in
which accelerators do the interaction calculation only, and CPUs do
all the rest. However, we try to minimize the communication between
CPUs and accelerators as much as possible, without making the
calculation on the side of accelerators very complicated.

In this paper, we describe our strategy of using accelerators, how
application programmers can use FDPS to efficiently use accelerators,
and achieved performance.

This paper is organized as follows. In section 2, we present the
overview of FDPS. In section 3, we discuss the traditional approach of
using accelerators for interaction calculation, and its
limitations. In section 4 we summarize our new approach. In section 5
we show how the users can use new APIs of FDPS to make use of
accelerators. In section 6 we give the result of performance
measurement for GPGPU-based systems and give the performance
prediction for hypothetical systems. Section 7 is for summary and
discussion.

\section{Overview of FDPS}

The basic idea (or the ultimate goal) of FDPS is to make it possible
for researchers to develop their own high-performance, highly-parallel
particle-based simulation codes without spending too much time for
writing, debugging, and performance tuning of the codes. In order to
achieve this goal, we have designed FDPS so that it provides all
necessary functions for efficient parallel program for particle-based
simulations. FDPS uses MPI for inter-node parallelization and OpenMP
for intra-node parallelization.  In order to reduce the communication
between computing nodes, the computational domain is divided using the
recursive multisection algorithm \citep{2004PASJ...56..521M}, but with
weights for particles to achieve the optimal load balancing
\citep{2009PASJ...61.1319I}. The number of subdomains is equal to the
number of MPI processes, and one subdomain is assigned to one MPI
process.

Initially, particles are distributed to MPI processes in an arbitrary
way. It is not necessary that the initial distribution is based on
spatial decomposition, and it is even possible that initially just one
process has all particles, if it has the sufficient amount of memory.
After the spatial coordinates of subdomains are determined, for each
particle, the MPI process to which it belongs is determined, and it is
sent to that process. These parts can be achieved just by calling FDPS
library functions. In order for FDPS functions to get information of
particles and copy or move them, FDPS functions need to know the data
structure of the particles. This is made possible by making FDPS
``template-based'', so that at compile time FDPS library functions
know the data structure of particles.

After particles are moved to their new locations, the interaction
calculation is done through parallel Barnes-Hut algorithm based on the
local essential tree \citep{2004PASJ...56..521M}.  In this method,
each MPI process first constructs the tree structure from its local
particles (local tree). Then, it sends, to all other MPI process, its
information necessary for that MPI process to evaluate the interaction
with its particles. This necessary information is called the local
essential tree (LET).

After one process received all LETs from all other nodes, it
constructs the global tree by merging the LETs. In FDPS, merging is
not actually done but LETs are first reduced to arrays of particles
and superparticles (hereafter we call it ``SPJ''), and a new tree is
constructed from combined list of all particles. Here, a SPJ
represents a node of Barnes-Hut tree.

Finally, the interaction calculation is done by traversing the tree
for each particles. Using Barnes' vectorization algorithm
\citep{1990JCoPh..87..161B}, we traverse the tree for a group of local
particles, and create the ``interaction list'' for that group. Then,
FDPS calculates the interaction exerted from particles and
superparticles in this interaction list to particles in the group, by
calling used-supplied interaction function.

In the case of long-range interaction, we use the standard Barnes-Hut
scheme for treewalk. In the case of short-range interaction such as
SPH interaction between particles, we still use treewalk but with
cell-opening criterion different from the standard opening angle.

Thus, users of FDPS can use the functions for domain decomposition,
particle migration and interaction calculation, by passing their own
particle data class and interaction calculation function to FDPS at
the compile time. Interaction calculation function should be designed
as receiving two arrays of particles, one exerting the ``force'' from
to the other.

\section{Traditional Approach to Use Accelerators and Its Limitation}

As we have already stated in the introduction, accelerators have been
used for gravitational $N$-body simulations, both on single and
parallel machines, with and without Barnes-Hut treecode
\citep{BarnesHut1986}.  In the case of the tree algorithm, the idea is
to use Barnes' vectorization algorithm, which is what we defined as
the interface between the user-defined interaction function and
FDPS. Thus, in principle we can use accelerators just by replacing the
user-defined interaction function with that uses the accelerators. In
the case of GRAPE processors, that would be the only thing we need to
do. At the same time, this would be the only thing we can do.

On modern GPGPUs, however, we need to modify the interface and
algorithm slightly. There are two main reasons for this
modification. The first one is that the software overhead of GPGPUs
for data transfer and kernel startup is much larger than that for
GRAPE processors. Another difference is in the architecture.  GRAPE
processors consist of relatively small number of highly pipelined,
application-specific pipeline processor for interaction calculation,
with hardware support for fast summation of results from multiple
pipelines. On the other hand, GPGPUs consist of a very large number of
programmable processors, with no hardware support for summation of the
results obtained on multiple processors. Thus, to make efficient use
of GPGPUs, we need to calculate interactions on a large number of
particles by single call to GPGPU computing kernel.  Vectorization
algorithm has one adjustable parameter, $n_{\rm grp}$, the number of
particles which share one interaction list, and it is possible to make
efficient use of GPGPUs by making this $n_{\rm grp}$ large. However,
using excessively large $n_{\rm grp}$ causes the increase of the total
calculation cost, and thus not desirable.

\citet{Hamadaetal2009} introduced an efficient way to use GPGPUs which
they called the ``mutltiwalk'' method. In their method, the CPU first
constructs multiple interaction lists for multiple groups of
particles, and then sends them to the GPGPU in a single kernel
call. GPGPU performs the calculation of multiple interaction lists in
parallel, and returns all results in a single data transfer. In this
way, we can tolerate the large overhead of invoking computing kernels
on GPGPUs and the lack of the support for fast summation.

Even though this multiwalk method is quite effective, there still
remain rooms of improvements, and that means on modern accelerators
the efficiency we can achieve with the mltiwalk method is rather
limited.

The biggest remaining inefficiency comes from the fact that with the
multiwalk method we send interaction lists for each particle group.
One interaction list is an array of physical quantities (at least
positions and masses) of particles. Typically, the number of particles
in an interaction list is ~10 times more than the number of particles
for which that interaction list is constructed, and thus the transfer
time of the interaction list is around 10 times longer than that of
the particles which receive the force. This means that we are sending
same particles (and superparticles) multiple times when we send
multiple interaction lists.

In the next section, we discuss how we can reduce the amount of
communication and also further reduce the calculation cost for the
parts other than the force calculation kernel.

\section{New Algorithms}
\label{ref:new_alg}

As we described in the previous section, to send all particles in the
interaction list to accelerators is inefficient because we send same
particles and SPJs multiple times. In section~\ref{ref:new_alg_id} we
will describe new algorithm to overcome this inefficiency. In
section~\ref{ref:new_alg_reuse}, we will also describe new algorithm
to further reduce the calculation cost for the parts other than the
force calculation kernel. In section~\ref{ref:new_alg_pro}, we will
describe the actual procedures with and without new algorithms.

\subsection{Indirect addressing of particles}
\label{ref:new_alg_id}

When we use the interaction list method on systems with accelerators,
in the simplest implementation, for each group of particles and its
interaction list, we send physical quantities necessary for
interaction calculation, such as positions and masses in the case of
gravitational force calculation. Roughly speaking, the number of
particles in the interaction list is around ten times longer than that
in one group. Thus, we are sending around $10n$ particles, where $n$
is the number of particles per MPI process, at each timestep.  Since
there are only $n$ local particles and the number of particles and
tree nodes in LETs is generally much smaller than $n$, this means that
we are sending the same data many times, and that we should be able to
reduce the communication by sending particle and tree node data only
once. Some GRAPE processors including GRAPE-2A, MDGRAPE-x and GRAPE-8,
have hardware support for this indirect addressing \citep{makino2012}.

In the case of programmable accelerators, this indirect addressing can
be achieved by first sending arrays of particles and tree nodes, and
then sending the interaction list (here the indices of particles and
tree nodes indicating the location of them in their arrays). The
user-defined interaction calculation function should be modified so
that it uses indirect addressing to access particles. Examples of such
code is included in the current FDPS distribution (version 4.0 and
later), and we plan to develop template routines which can be used to
generate codes on multiple platforms from single user-supplied code
for interaction calculation.

Interaction list is usually an array of 32-bit integers (four bytes),
and one particle data is at least 16 bytes (when positions and masses
are all in single precision numbers), but can be much larger in the
case of SPH and other method. Thus, with this method we can reduce the
communication cost by a large factor.

One limitation of this indirect addressing method is that all
particles in one process should fit in the memory of the accelerator.
Most of accelerator have relatively small memories. In such cases, we
can still use this method, by dividing the particles into blocks small
enough to fit the memory of accelerator. For each block, we construct
the ``global'' tree structure similar to that for all particles in the
process, and interaction lists for all groups under the block.

\subsection{Reuse of Interaction Lists}
\label{ref:new_alg_reuse}

For both of long-range and short-range interactions, FDPS constructs
the interaction lists for groups of particles. It is possible to keep
using the same interaction lists for multiple timesteps, if particles
do not move large distances in a single timestep.  In the case of SPH
or molecular dynamics simulations, it is guaranteed that particles
move only a small fraction of interparticle distance in a single
timestep, since the size of the timestep is limited by the stability
condition. Thus, in such cases we can safely use the interaction lists
for several timesteps.

Even in the case of gravitational many-body simulations, there are
cases where the change of the relative distance of between particles
in a single timestep is small. For example, both in the simulations of
planetary formation processes or planetary rings, the random
velocities of particles are very small, and thus, even though
particles move large distances, there is no need to reconstruct the
tree structure at each timestep, because the changes of the relative
positions of particles are small.

In the case of galaxy formation simulation using Nbody+SPH technique,
generally the timestep for the SPH part is much smaller than that for
the gravity part, and thus we should be able to use the same tree
structure and interaction lists for multiple SPH steps.

If we use this algorithm (hereafter we call it the reuse algorithm),
the procedures of the interaction calculation for the step with tree
construction and that without the tree construction are different. The
procedure for the tree construction step is given by
\begin{enumerate}
\item Construct the local tree
\item Construct the LET for all other processes. These LETs  should be the
  list of indices of particles and tree nodes, so that they can be
  used later.
\item  Exchange LETs. Here, the physical information of tree nodes and
  particles should be exchanged
\item Construct the global tree
\item Construct the interaction lists
\item Perform the interaction calculation for each group using the
  constructed list.  
\end{enumerate}

The procedure for reusing steps is given by
\begin{enumerate}
\item Update the physical information of the local tree
\item Exchange LETs.
\item Update the physical information of the global tree
\item Perform the interaction calculation for each group using the
  constructed list.  
\end{enumerate}  

In many cases we can keep using the same interaction list for around
10 timesteps. In the case of planetary ring simulation, using the same
list for much larger number of timesteps is possible, because the
stepsize of planetary ring simulation using ``soft-sphere'' method
\citep{Iwasawaetal2018} is limited by the hardness of the ``soft''
particles and thus much smaller than the usual timescale determined by
the local velocity dispersion and interparticle distance.

With this reuse algorithm, we can reduce the cost of the following
steps: (a) tree construction, (b) LET construction, (c) interaction
list construction. The calculation costs of steps (a) and (c) are
$O(N)$ and $O(N \log N)$, respectively. Thus they are rather large for
simulations with large number of particles. Moreover, by reducing the
cost of step (c), we can make the group size $n_{\rm grp}$ small,
which results in the decrease of the calculation cost due to the use
of interaction list. Thus, the overall improvement of the efficiency
is quite significant.

The construction and maintenance of interaction lists and other
necessary data structures are all done within FDPS. Therefore,
user-developed application programs can use this reuse algorithm just
by calling the FDPS interaction calculation function with one
additional argument indicating reuse/construction. The necessary
change of the application program is very small.

\subsection{Procedures with or without New Algorithms}
\label{ref:new_alg_pro}

In this section, we describe the actual procedures of simulations
using FDPS with or without new algorithms. Before describing the
procedures, let us introduce four particle data types FDPS uses: FP
(Full Particle), EPI (Essential Particle I), EPJ (Essential Particle
J) and FORCE.  FP is the data structure containing all information of
a particle, EPI(J) is used for the minimal data of particles which
receives (gives) the force, and FORCE type to store the calculated
interaction. FDPS uses these additional three data types to minimize
the memory access during the interaction calculation. We first
describe the procedure for the calculation without the reuse algorithm
and then describe that for the reuse algorithm.

At the beginning of one timestep, the computational domains assigned
to MPI processes are determined and all processes exchange particles
so that all particles belong to their appropriate domains. Then, the
coordinates of the root cell of the tree are determined using the
positions of all particles. After the determination of the root cell,
each MPI process constructs its local tree. The local tree
construction consists of the following four steps.
\begin{enumerate}
\item Generate Morton keys for all particles.
\item Sort key-index pairs in Morton order by the radix sort.
\item Reorder FPs in Morton order referring the key-index pairs and
  copy the particle data from FPs to EPIs and EPJs.
\item For each level of the tree, from top to bottom, allocate tree
  cells and link their child cells. In each level, we use the binary
  search to find cell boundaries.
\end{enumerate}
In the case of the reusing step, these steps are skipped.

After the construction of the local tree, multipole moments of all
local tree cells are calculated, from the bottom to the top of the
tree. Even at the reusing step, the calculation of the multipole
moments is performed because the physical quantities of particles are
updated at every timesteps.

After the calculation of the multipole moments of the local tree, each
MPI process constructs LETs and send them to other MPI process. When
the reusing algorithms is used, at the tree construction step, each
MPI process saves the LETs and their destination processes.

After the exchange of LETs, each MPI process constructs the global
tree from received LETs and its local tree. The procedure is almost
the same as that for the local tree construction.

After the construction of the global tree, each MPI process calculates
the multipole moments of all cells of the global tree. The procedure
is the same as that for the local tree.

After the calculation of the moments of the global tree, each MPI
process constructs the interaction lists and using them performs the
force calculation. If we do not use the multiwalk method, each MPI
process makes the interaction lists for one particle group and then
the user-defined force kernel calculates the forces from EPJs and SPJs
in the interaction list to EPIs in the particle group.

When we use the multiwalk method, each MPI process makes multiple
interaction lists for multiple particle groups. When the indirect
addressing method is not used, each MPI process gives multiple groups
and multiple interaction lists to the interaction kernel on the
accelerator. Thus we can summarize the procedure of the force
calculation without the indirect addressing method for multiple
particle groups as follows:
\begin{enumerate}
\item Construct the interaction list for multiple particle groups.
\item Copy EPIs and the interaction lists to the send buffer for the
  accelerator. Here, the interaction list consists of EPJs and SPJs.
\item Send particle groups and their interaction lists to the accelerator.
\item Let the accelerator calculate interactions on particle groups
  sent at step 3.
\item Receive the results calculated at step 4 and copy them back to
  FPs, integrate orbits of FPs and copy the data from FPs to EPIs and
  EPJs.
\end{enumerate}

To calculate forces on all particles, the above steps are repeated
until all particle groups are processed. Note that the construction of
the interaction list (step 1), sending the data to the accelerator
(step 3), actual calculation (step 4) and receiving the calculated
result (step 5) can all be overlapped.

On the other hand, when the indirect addressing method is used, before
the construction of the interaction lists, each MPI process sends the
data of all cells of the global tree to the accelerator. Thus at the
beginning of the interaction calculation, it should send them to the
accelerator. After that, the accelerator receives the data of particle
groups and their interaction lists but here the interaction list
contains the indices of EPJs and SPJs and not their physical
quantities.  Thus, the calculation procedure with indirect addressing
method is the same as that without the indirect addressing except that
all data of the global tree are sent at the beginning of the
calculation and the interaction lists sent during the calculation
contains only indices of tree cells and EPJs.

Both with and without the indirect addressing method, we can use the
reusing method. For the construction step, the procedures are the
same. For the reusing steps, we can skip the steps for the
interaction-list construction (step 1).  When we use the indirect
addressing method, we can also skip the sending of them since the
lists of indices are unchanged during the reuse.

\section{APIs to use Accelerators}
\label{sec:api}

In this section, we describe the APIs (application program interfaces)
of FDPS to use accelerators and how to use them by showing sample
codes developed for NVIDIA GPGPUs. Part of the user kernel is written
in CUDA.

FDPS has high level APIs to perform all procedures for interaction
calculation in single API call. For the multiwalk method, FDPS
provides {\tt calcForceAllAndWriteBackMultiWalk} or {\tt
  calcForceAllAndWriteBackMultiWalkIndex}. The difference between
these two functions is that the former dose not use the indirect
addressing method. These two APIs can be used as the replacement of
{\tt calcForceAllAndWriteBack}, which is another top level API
provided by FDPS distribution version 1.0 or later. A user must
provide two force kernels: the ``dispatch'' and ``retrieve''
kernels. The ``dispatch'' kernel is used to send EPIs, EPJs and SPJs
to accelerators and call the force kernel. The ``retrieve'' kernel is
used to collect FORCEs from accelerators. The reason why FDPS needs
two kernels is to allow the overlap of the calculation on the CPU with
the force calculation on the accelerator as we described in the
previous section.

\begin{figure}
  \begin{center}
    \includegraphics[width=10cm]{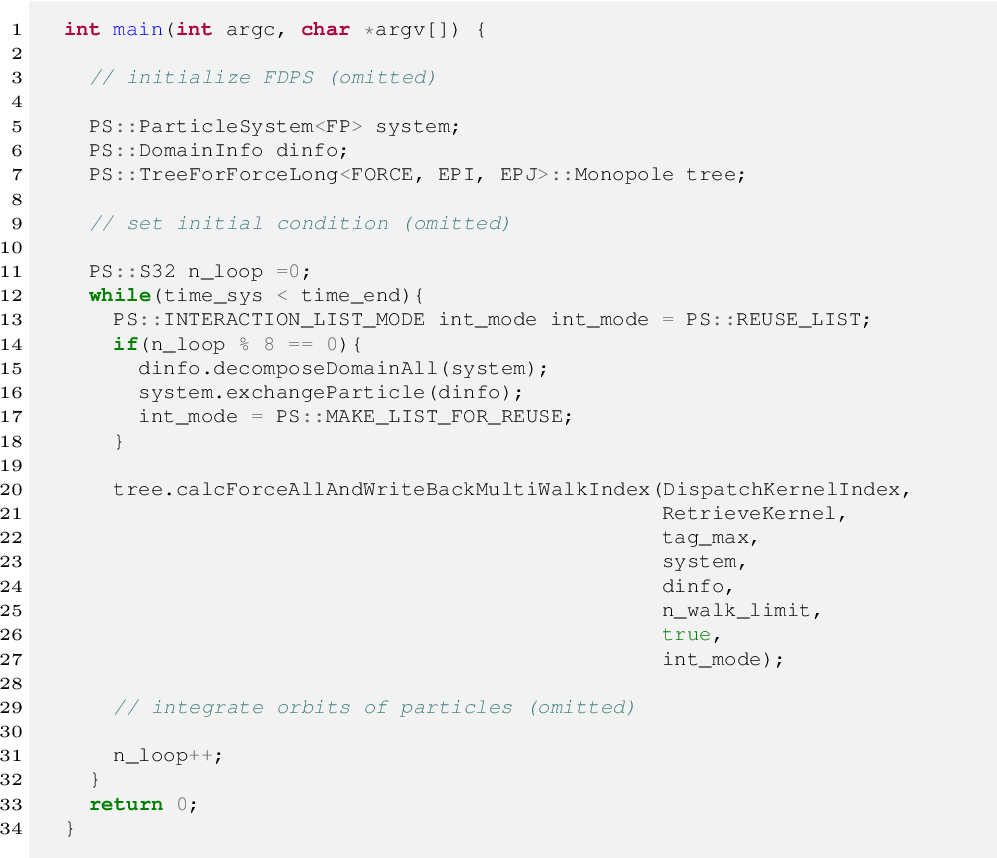}
    \end{center}
    \caption{Example of how to use the reusing method.}
  \label{fig:reuse}
\end{figure}

The reusing method can be used with all of three top level APIs
described above. The only thing users do to use the reusing method is
to give an appropriate FDPS-provided enum-type value to these
functions so that the reusing method is enabled. The enum-type values
FDPS provided are {\tt MAKE\_LIST}, {\tt MAKE\_LIST\_FOR\_REUSE} and
{\tt REUSE\_LIST}. At the construction step the application program
should give {\tt MAKE\_LIST\_FOR\_REUSE} to the top level APIs so that
FDPS constructs the trees and the interaction lists and saves them. At
the reusing step, the application program should give {\tt
  REUSE\_LIST} so that FDPS skips the construction of the trees and
reuses the interaction lists constructed at the last construction
step. In the case of {\tt MAKE\_LIST}, FDPS also constructs the trees
and the interaction lists but dose not save them. Thus the users
cannot use the reusing method. Figure~\ref{fig:reuse} shows an example
of how to use the reusing method. In this example, the trees and the
interaction lists are constructed once in every eight steps. While the
same list is being reused, particles should remain in the same MPI
process as at the moment of the list construction. Thus {\tt
  exchangeParticle} should be called only just before the tree
construction step.

\begin{figure}
  \begin{center}
    \includegraphics[width=18cm]{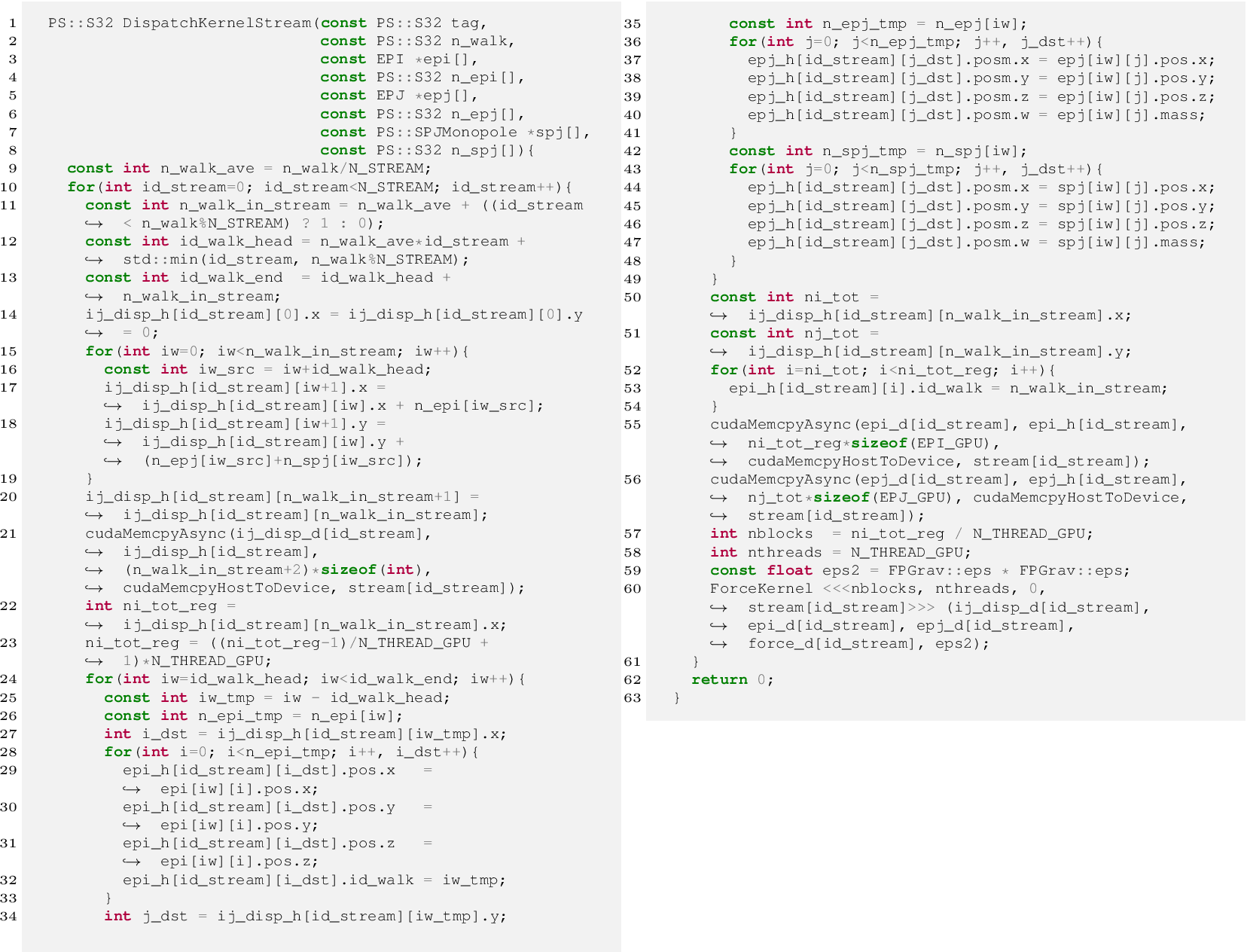}
    \end{center}
    \caption{Example of the dispatch Kernel without the indirect addressing method.}
  \label{fig:dispatch}
\end{figure}

Figure~\ref{fig:dispatch} shows an example of the dispatch kernel
without the indirect addressing method. FDPS gives the dispatch kernel
the arrays of the pointers of EPIs, EPJs and SPJs as the arguments of
the kernel (lines 3, 5 and 7). Each pointers point to the address of
the first elements of the arrays of EPIs, EPJs and SPJs for one group
and its interaction list.  The sizes of these arrays are given by
${\rm n\_epi}$ (line 4), ${\rm n\_epj}$ (line 6) and ${\rm n\_spj}$
(line 8). FDPS also gives ``tag'' (the first argument) and ``n\_walk''
(the second argument). The argument ``tag'' is used to specify
individual accelerators if multiple accelerators are
available. However, in the current version of FDPS, ``tag'' is
disabled and FDPS always gives it 0. The argument ``n\_walk'' is the
number of the particle groups and interaction lists.

To overlap the actual force calculation on the GPGPU with the data
transfer between the GPGPU and the CPU, we use CUDA stream, which is a
sequence of operations executed on the GPGPU. In this example, we used
{\tt N\_STREAM} CUDA streams. In this paper, we used 8 CUDA streams
because even if we use more CUDA streams, the performance of our
simulations is not improved. In each stream, ${\rm n\_walk /
  N\_STREAM}$ interaction lists are handled. The particle data types,
EPIs (lines 28--33), EPJs (lines 36--41) and SPJs (lines 48--48) are
copied to the send buffers for GPGPUs. Here, we use the same buffer
for EPJs and SPJs because the types of the EPJ and SPJ are the
same. In lines 55 and 56, the EPIs and EPJs are sent to the
GPGPU. Then the force kernel is called in line 60.

\begin{figure}
  \begin{center}
    \includegraphics[width=18cm]{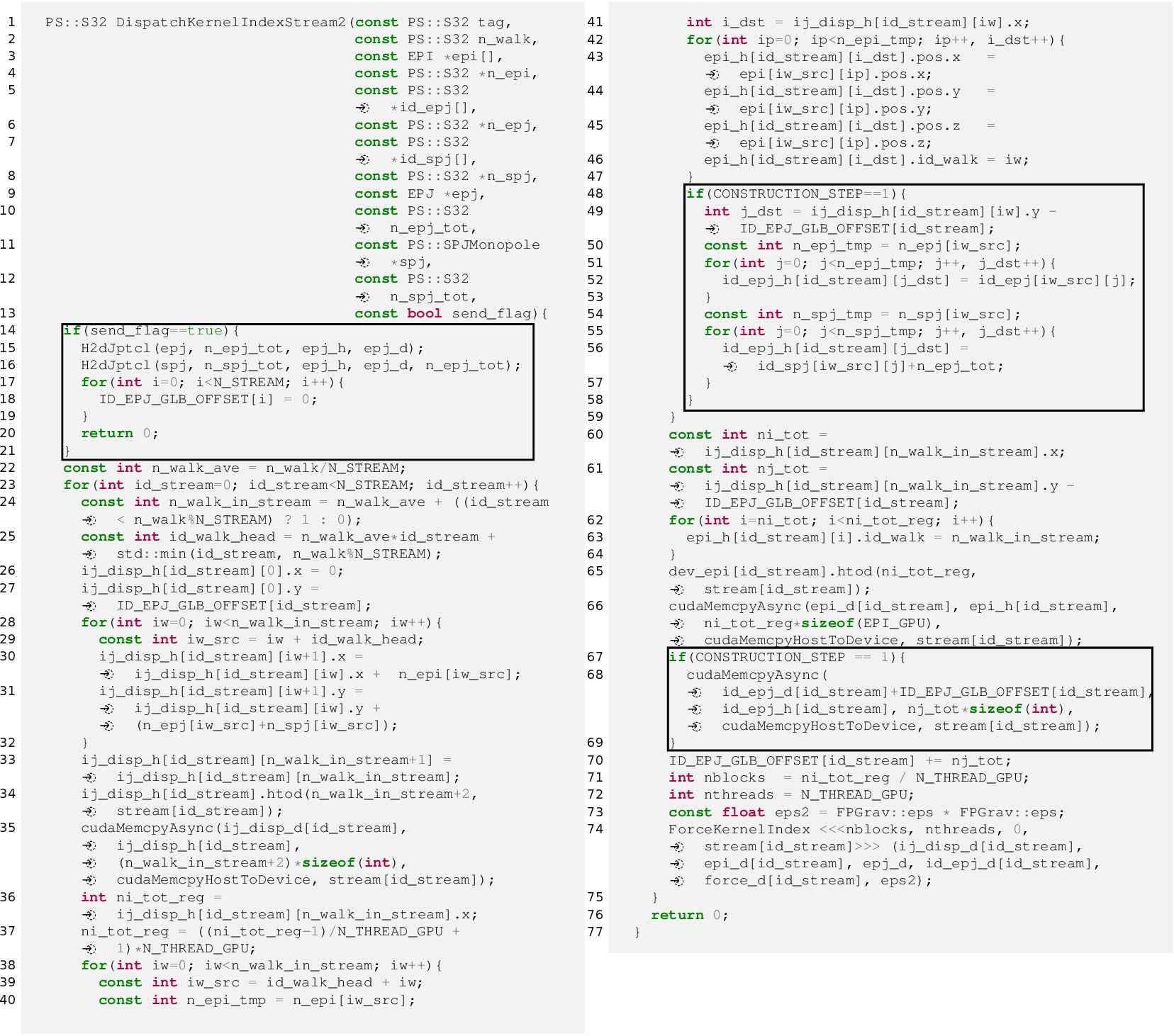}
    \end{center}
    \caption{Example of the dispatch Kernel with the indirect
      addressing method. Boxes with solid line indicate the
      differences from the kernel without indirect addressing
      method. }
  \label{fig:dispatch_index}
\end{figure}

Figure~\ref{fig:dispatch_index} shows an example of the dispatch
kernel with the indirect addressing method. This kernel is almost the
same as that without the indirect addressing method except for two
differences. One difference is that at the beginning of one timestep,
all data of the global tree (EPJs and SPJs) are sent to GPGPU (lines
15 and 16). Whether or not the application program sends EPJs and SPJs
to GPGPU is specified by the 13th argument ``send\_flag''. If
``send\_flag'' is true, the application program sends all EPJs and
SPJs. Another difference is that indices of EPJs and SPJs are sent
(lines 48--58 and 67--69) instead of physical quantities of EPJs and
SPJs. Here, we use user-defined global variable {\tt
  CONSTRUCTION\_STEP} to specify whether the current step is
construction or reusing steps. At the construction step, {\tt
  CONSTRUCTION\_STEP} becomes unity and the user program sends the
interaction list to the GPGPU and saves them in the GPGPU. On the
other hand, at the reusing step, the user program dose not send the
list and reuse the interaction list saved in the GPGPU.

\begin{figure}
  \begin{center}
    \includegraphics[width=10cm]{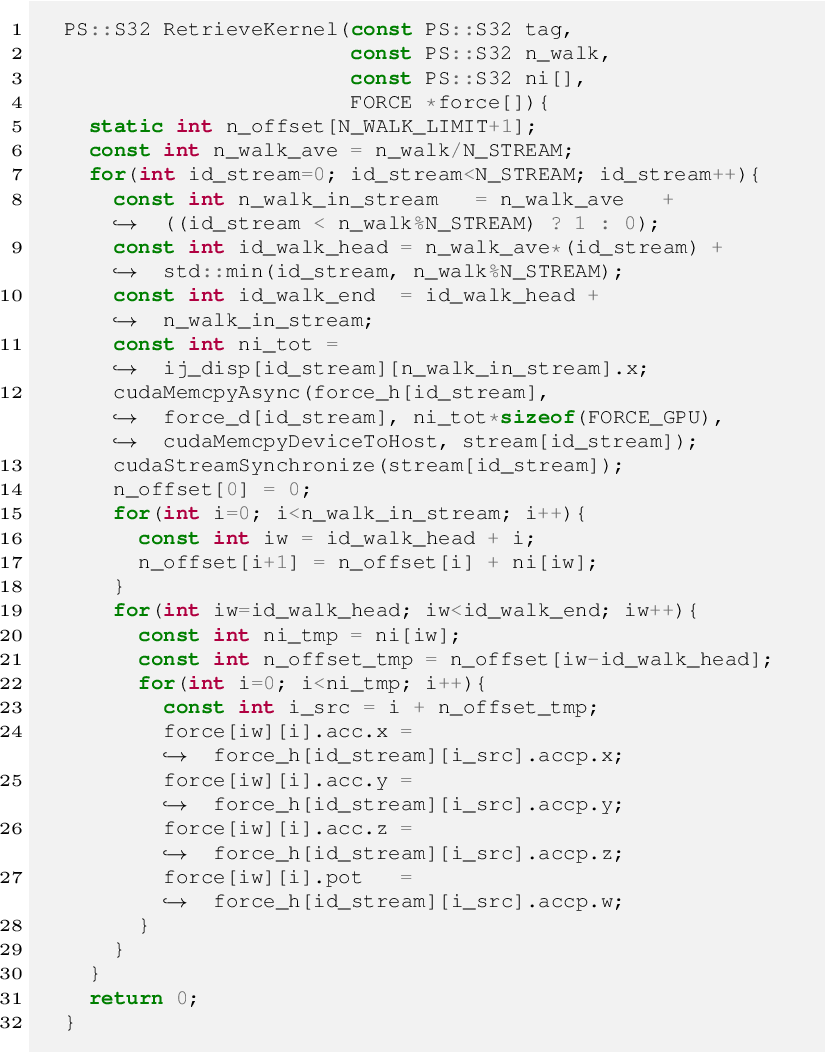}
    \end{center}
    \caption{Example of the retrieve kernel.}
  \label{fig:retrieve}
\end{figure}

Figure~\ref{fig:retrieve} shows an example of the retrieve kernel. The
same retrieve kernel can be used with and without the indirect
addressing method. In line 12, the GPGPU sends the interaction results
to the receive buffer of the host. To let the CPU wait until all
functions in the same stream on the GPGPU are completed, {\tt
  cudaStreamSynchronize} is called in line 13. Finally, the
interaction results are copied to FORCEs.

\section{Performance}
\label{sect:performance}

In this section, we present the measured performance and the
performance model of a simple $N$-body simulation code implemented
using FDPS on CPU-only and CPU+GPGPU systems.

\subsection{Measured Performance}
\label{sect:measuredperformance}

To measure the performance of FDPS, we performed simple gravitational
$N$-body simulations both with and without the accelerator. The
initial model is a cold uniform sphere. This system will collapse in a
self-similar way. Thus we can use the reusing method. The number of
particles (per process) $n$ is $2^{22}$ (4M). The opening criterion of
the tree, $\theta$, is between 0.25 and 1.0, the maximum number of
particles in a leaf cell is 16 and the maximum number of particles in
EPI group, ${n_{\rm grp}}$, is between 64 and 16384. We performed
these simulations with three different methods. We listed all methods
in table~\ref{tab:all_method}. In the case of the reusing method, the
number of reusing steps between the construction steps ${n_{\rm
    reuse}}$ is between 2 and 16.

\begin{table}
  \begin{tabular}{ccccc}
    \toprule
     Method    & System   & Multiwalk  & Indirect Addressing & Reusing \\
     \midrule
     C1        & CPU        & No         & No               & No        \\
     G1        & CPU+GPGPU  & Yes        & No               & No        \\
     G2        & CPU+GPGPU  & Yes        & Yes              & Yes       \\
  \bottomrule
  \end{tabular}
  \caption{All methods we used to perform simulations.}
  \label{tab:all_method}
\end{table}

We used NVIDIA TITAN V as an accelerator. Its peak performance is 13.8
Tflops for single precision calculation. Host CPU is single Intel Xeon
E5-2670 v3 with the peak speed of 883 Gflops for single precision
calculation. The GPGPU is connected to the host CPU through PCI
Express 3.0 bus with 16 lanes. The main memory of the host computer is
DDR4-2133. The theoretical peak bandwidth is 68 GB/s for the host main
memory and 15.75 GB/s for the data transfer between the host and
GPGPU. All data of particles are in double precision. Force
calculation on GPGPU and data transfer between the CPU and GPGPU are
performed in single precision.

\begin{figure}
    \begin{center}
      \includegraphics[width=12cm]{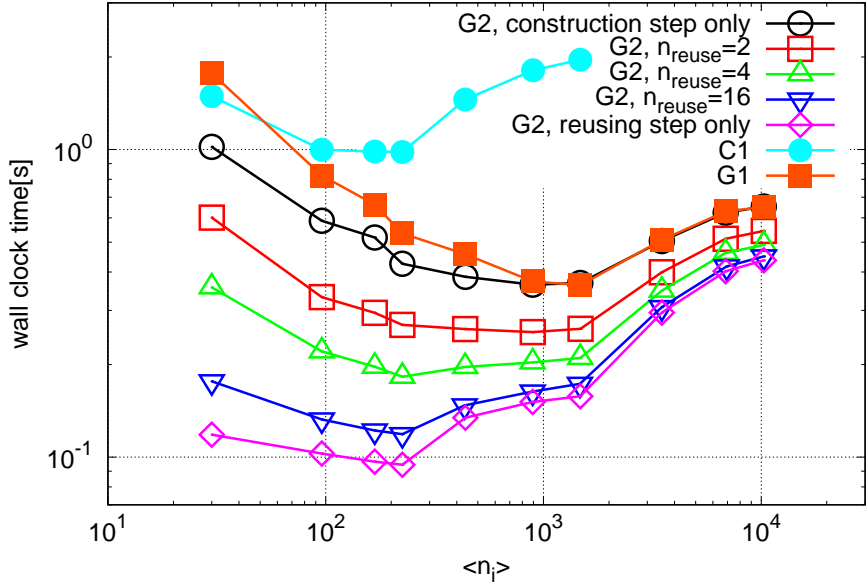}
    \end{center}
    \caption{Averaged elapsed time per step against $ \langle n_{\rm
        i} \rangle $. Open symbols indicate the results of the runs
      with method G2. Open squares, open upward triangles and open
      inverted triangles indicate the results with $n_{\rm reuse}$ of
      2, 4 and 16, respectively. Open circles and open diamonds
      indicate the elapsed times of the construction step and the
      reusing step, respectively. Filled circles and filled squares
      indicate the elapsed times for methods C1 and G1, respectively.
    }
  \label{fig:wtime_ngrp_th0.5}
\end{figure}

Figure~\ref{fig:wtime_ngrp_th0.5} shows the average elapsed time per
step for methods C1, G1 and G2 with the reuse interval $n_{\rm reuse}$
of 2, 4 and 16 plotted against the average number of particles which
share one interaction list $\langle n_{\rm i} \rangle$. The opening
angle is $\theta=0.5$. We also plot the elapsed times for method G2 at
construction step and at reusing step. The difference between the
elapsed time for method G1 and that for G2 at the construction step is
due to the difference in the use of the indirect addressing method.

We can see that, in the case of method G2, the elapsed time becomes
smaller as reuse interval $n_{\rm reuse}$ increases, and approaches to
the time of the reusing step. The minimum time of method G2 at the
reusing step is ten times smaller than that of method C1 and four
times smaller than that of method G1. The performance of method G2
with $n_{\rm reuse}$ of 16 is 3.7 Tflops (27 \% of the theoretical
peak).

We can also see that the optimal values of $\langle n_{\rm i} \rangle$
becomes smaller as $n_{\rm reuse}$ increases. When we do not use the
reuse method, the tree construction and traversal are done at every
step. Thus, their costs are large, and to make it small we should
increase $\langle n_{\rm i} \rangle$. In order to do so, we need to
use large $n_{\rm crit}$, which is the maximum number of particles in
the particle group. If we make $\langle n_{\rm i} \rangle$ too large,
the calculation cost increases \citep{1991PASJ...43..621M}. Thus there
is an optimal $\langle n_{\rm i} \rangle$. When we use the reuse
method, the relative cost of tree traversal becomes smaller. Thus, the
optimal $\langle n_{\rm i} \rangle$ becomes smaller and the
calculation cost also becomes smaller. We will give more detailed
analysis in section~\ref{sect:performancemodel}.

Table~\ref{tab:time} shows the breakdown of the calculation time for
different methods in the case of $\theta$ of 0.5. For the runs with C1
and G1, we show the breakdown at the optimal values of $\langle n_{\rm
  i} \rangle $, which are 230 and 1500, respectively. For the run with
G2, we show the breakdowns of the construction and the reusing steps
for $\langle n_{\rm i} \rangle $ of 230. We can see that the
calculation time for reusing step is four times smaller than that for
construction step. Thus if $n_{\rm reuse} \gg 4$, the contribution of
the construction step to the total calculation time is small.

\begin{table}
  \begin{tabular}{lccccc}
    \toprule
    method  & C1 & G1  & G2 (construction step) & G2 (reusing step) \\
    \midrule
    $\langle n_{\rm i} \rangle$ & 230 & 1500 & 230 & 230 \\
    \midrule
    set root cell                                     & 0.0064  & 0.0066   & 0.0066   & - \\
    make local tree                                   & 0.091   & 0.092    & 0.095    & - \\
    \hspace{8dd} calculate key                        & 0.0084  & 0.0085   & 0.0093   & - \\
    \hspace{8dd} sort key                             & 0.042   & 0.043    & 0.044    & - \\
    \hspace{8dd} reorder ptcl                         & 0.030   & 0.030    & 0.030    & - \\
    \hspace{8dd} link tree cell                       & 0.011   & 0.010    & 0.011    & - \\
    calculate multipole moment of local tree          & 0.0053  & 0.0058   & 0.0053   & 0.0062 \\
    make global tree                                  & 0.094   & 0.094    & 0.096    & 0.0071  \\
    \hspace{8dd} calculate key                        & 0.0     & 0.0      & 0.0      & -      \\
    \hspace{8dd} sort key                             & 0.040   & 0.041    & 0.042    & -      \\
    \hspace{8dd} reorder ptcl                         & 0.036   & 0.036    & 0.037    & 0.0071 \\
    \hspace{8dd} link tree cell                       & 0.011   & 0.011    & 0.011    & -      \\
    calculate multipole moment of global tree         & 0.0066  & 0.0064   & 0.0065   & 0.0068  \\
    calculate force                                   & 0.76    & 0.15     & 0.21     & 0.072   \\
    \hspace{8dd} make interaction list (EPJ and SPJ)  & (0.16)  & 0.063    & -        & -        \\
    \hspace{8dd} make interaction list (id)           & -       & -        & 0.13    & -        \\
    \hspace{8dd} copy all particles and tree cells    & -       & -        & (0.0065) & (0.0065) \\ 
    \hspace{8dd} copy EPI                             & -       & (0.0037) & (0.0037) & (0.0037) \\ 
    \hspace{8dd} copy interaction list (EPJ and SPJ)  & -       & (0.020)  & -        & -        \\ 
    \hspace{8dd} copy interaction list (id)           & -       & -        & (0.013)  & -        \\ 
    \hspace{8dd} copy FORCES                          & -       & (0.0060) & (0.0060) & (0.0060) \\ 
    \hspace{8dd} force kernel                         & (0.43)  & (0.11)   & (0.043)  & (0.043)  \\
    \hspace{8dd} H2D all particles and tree cells     & -       & -        & (0.0073) & (0.0073) \\ 
    \hspace{8dd} H2D EPI                              & -       & (0.0042) & (0.0042) & (0.0042)  \\  
    \hspace{8dd} H2D interaction list (EPJ and SPJ)   & -       & (0.034)  & -        & -         \\  
    \hspace{8dd} H2D interaction list (id)            & -       & -        & (0.022)  & -         \\  
    \hspace{8dd} D2H FORCE                            & -       & (0.0067) & (0.0067) & (0.0067)  \\
    \hspace{8dd} writ back + integration (+ copy ptcl)& 0.015   & 0.016    & 0.021    & 0.025  \\
    \midrule
    total                                             & 0.98    & 0.36     & 0.42     & 0.095 \\
    \midrule
  \bottomrule
  \end{tabular}
  \caption{Breakdown of the total time per step for the runs with
    various methods in the case of $\theta$ of 0.5. The second row
    shows $\langle n_{\rm i} \rangle$ we used in this measurement. The
    times in parentheses are the estimated times by using the
    performance model we will discuss in the following sections,
    because the calculations corresponding to these times are hidden
    by other calculation and we can not measure them.}
  \label{tab:time}
\end{table}

\begin{figure}
    \begin{center}
      \includegraphics[width=12cm]{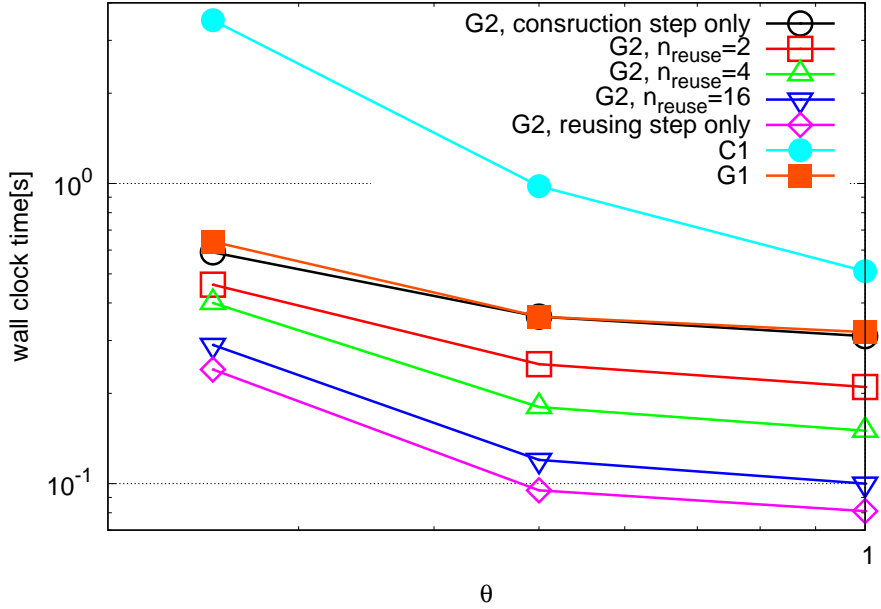}
    \end{center}
    \caption{Total times for various methods at the optimal $\langle
      n_{\rm i} \rangle$ against $\theta$. Symbols are the same as in
      figure~\ref{fig:wtime_ngrp_th0.5}.}
  \label{fig:theta_wtime}
\end{figure}

Figure~\ref{fig:theta_wtime} shows the average elapsed time at optimal
$\langle n_{\rm i} \rangle$ plotted against $\theta$ for methods C1,
G1 and G2 with $n_{\rm reuse}$ of 2, 4 and 16. We also plot the
elapsed times for construction and reusing steps of method G2. The
slope for method C1 is steeper than those for other methods. This is
because the time for the force kernel dominates the total time in the
case of method C1 and it strongly depends on $\theta$.

\subsection{Performance Model on Single Node}
\label{sect:performancemodel}

In the following, we present the performance model of the $N$-body
simulation with FDPS with the monopole approximation on a single node
with and without accelerator. The total execution times per step on
single node for construction step $T_{\rm step, const}$ and for
reusing step $T_{\rm step, reuse}$ are given by
\begin{eqnarray}
  T_{\rm step, const} & \sim & T_{\rm root} + T_{\rm const\, lt} + T_{\rm mom\, lt} + T_{\rm const\, gt} + T_{\rm mom\, gt} + T_{\rm force, const}, \label{eq:time_single0} \\
  T_{\rm step, reuse} & \sim & T_{\rm mom\, lt} + T^{\rm reuse}_{\rm reorder\, gt} + T_{\rm mom\, gt} + T_{\rm force, reuse} \label{eq:time_single1},
\end{eqnarray}
where $T_{\rm root}$, $T_{\rm const\, lt}$, $T_{\rm mom\, lt}$,
$T_{\rm const\, gt}$, $T_{\rm mom\, gt}$, $T_{\rm force, const}$,
$T^{\rm reuse}_{\rm reorder\, gt}$ and $T_{\rm force, reuse}$ are the
times for the determination of the root cell of the tree, the
construction of the local tree, the calculation of the multipole
moments of the cells of the local tree, the construction of the global
tree, the calculation of the multipole moments of the cells of the
global tree, the force calculation for the construction step, the
reorder of the particles for the global tree and the force calculation
for the reusing step. The force calculation times $T_{\rm force,
  const}$ and $T_{\rm force, reuse}$ include the times for the
construction of the interaction list, the actual calculation of the
interactions, the copy of the interaction results from FORCEs to FPs,
the integration of orbits of FPs and the copy of the particle data
from FPs to EPIs and EPJs. In the reuse step, the tree is reused and
therefore $T_{\rm root}$, $T_{\rm const\, lt}$ and $T_{\rm const\,
  gt}$ do not appear in $T_{\rm step, reuse}$. On the other hand,
$T^{\rm reuse}_{\rm reorder\, gt}$ appears in $T_{\rm step,
  reuse}$. This is because in the reusing step the particles are
sorted in Morton order of the local tree and the particles should be
reordered in Morton order of the global tree. In the following
subsections, we will discuss the elapsed times of each component in
equations~(\ref{eq:time_single0}) and (\ref{eq:time_single1}).

\subsubsection{Model of $T_{\rm root}$}

At the beginning of the construction step, the coordinate of the root
cell of the tree is determined. In order to do so, CPU cores read the
data of all particles (FPs) from the main memory, and on modern
machines the effective main memory bandwidth determines the
calculation time. The actual computation of determination of the root
cell coordinate is much faster compared to the main memory access.
Thus $T_{\rm root}$ is given by
\begin{equation}
  T_{\rm root} \sim \frac{n \alpha_{\rm root} b_{\rm FP}}{B_{\rm host}},
  \label{eq:model_root_cell}
\end{equation}
where $n$ is the number of local particles, $\alpha_{\rm root}$ is the
coefficient for the memory access speed, $b_{\rm FP}$ is the data size
of FP and $B_{\rm host}$ is the effective bandwidth of the main memory
of the host which was measured by {\tt STREAM} benchmark. Note that we
used the ``copy'' kernel of {\tt STREAM} benchmark. In other words,
the bandwidth $B_{\rm host}$ indicates the mean bandwidth of reading
and writing. On our system, the bandwidth of reading is slightly
faster than that of writing. Thus for reading-dominant procedure,
$\alpha$ would be smaller than unity. For our $N$-body code, we found
$\alpha_{\rm root} \sim 0.7$. The coefficients for
equation~(\ref{eq:model_root_cell}) are listed in
table~\ref{tab:model_root_cell}.

\begin{table}
  \caption{Coefficients for equation~(\ref{eq:model_root_cell})}
  \label{tab:model_root_cell}
  \begin{tabular}{ll}
    \toprule
    $\alpha_{\rm root}$  & 0.70 \\
    $b_{\rm FP}$  & 88 byte \\
  \bottomrule
  \end{tabular}
\end{table}

\subsubsection{Model of $T_{\rm const lt}$}

The time for the construction of the local tree $T_{\rm const\, lt}$
is given by
\begin{equation}
  T_{\rm const\, lt}  \sim T_{\rm key\, lt} + T_{\rm sort\, lt} + T_{\rm reorder\, lt} + T_{\rm link\, lt},
\end{equation}
where $T_{\rm key\, lt}$, $T_{\rm sort\, lt}$, $T_{\rm reorder\, lt}$
and $T_{\rm link\, lt}$ are the elapsed times for the calculation of
Morton keys, sorting of the key-index pairs, reordering of FPs, EPIs
and EPJs by using the sorted key-index pairs and linking of tree
cells.

The time for key construction is determined by the time to read FPs
and write key-index pairs. Thus $T_{\rm key\, lt}$ is given by
\begin{equation}
  T_{\rm key\, lt}   \sim  \frac{n \alpha_{\rm key} \left( b_{\rm FP}+b_{\rm key} \right)}{B_{\rm host}},
  \label{eq:lt_key}
\end{equation}
where $b_{\rm key}$ is the data size of the key-index pair.

For sorting, we use the radix sort \citep{knuth1997} with the chunk
size of eight bits for the keys with 64-bit length. Thus we need to
apply the basic procedure of the radix sort eight times. For each
chunk, the data are read twice and written once. Thus the total number
of memory access is 24. Therefore $T_{\rm sort\, lt}$ is given by
\begin{equation}
T_{\rm sort\, lt}  \sim n \frac{\alpha_{\rm sort} 24 b_{\rm key}}{B_{\rm host}}.
\end{equation}

For reordering, the key-index pair and FP are read once and FP, EPI
and EPJ are written and 
\begin{equation}
T_{\rm reorder\, lt} \sim n \frac{\alpha_{\rm reorder\, lt} \left(2b_{\rm FP}+b_{\rm EPI}+b_{\rm EPJ}+b_{\rm key} \right)}{B_{\rm host}},
\end{equation}
where the size parameters $b_{\rm EPI}$ and $b_{\rm EPJ}$ are those of
the EPI and EPJ, respectively.

In the linking part, we generate the tree structure from the sorted
keys.  In order to do so, for each cell of the tree in each level, we
determine the location of the first particle by the binary search
method. Thus the cost is proportional to $n_{\rm cell} \log_2\left(
n_{\rm cell}\right)$ where $n_{\rm cell}$ is the total number of tree
cells. For usual Barnes-Hut tree, we have $n_{\rm cell} \sim n/4$.
Thus $T_{\rm link\, lt}$ is given by
\begin{equation}
  T_{\rm link\, lt} \sim \frac{n}{4} \log_2{ \left( \frac{n}{4} \right) } \frac{\alpha_{\rm link} b_{\rm key} }{B_{\rm host}}.
  \label{eq:lt_link}
\end{equation}

The size parameters $b$ and the memory access efficiency parameters
$\alpha$ in equations~(\ref{eq:lt_key}) to (\ref{eq:lt_link}) are
listed in table~\ref{tab:model_lt}. The reason why $\alpha_{\rm link\,
  lt}$ is much larger than unity is that in the binary search the
address to access depends on the data just read.

\begin{table}
  \caption{Coefficient for equations~(\ref{eq:lt_key}) to (\ref{eq:lt_link})}
  \label{tab:model_lt}
  \begin{tabular}{ll}
    \toprule
    $\alpha_{\rm key}$         & 0.85 \\
    $\alpha_{\rm sort}$        & 1.1  \\
    $\alpha_{\rm reorder\, lt}$  & 1.1 \\
    $\alpha_{\rm link}$        & 3.4 \\
    $b_{\rm key}$ & 16 byte \\
    $b_{\rm EPI}$ & 24 byte \\
    $b_{\rm EPJ}$ & 32 byte \\
  \bottomrule
  \end{tabular}
\end{table}

\subsubsection{Model of $T_{\rm mom\, lt}$}

The time for the calculation of the multipole moments of the local
tree is determined by the time to read EPJ and therefore $T_{\rm mom\,
  lt}$ is given by
\begin{equation}
  T_{\rm mom\, lt} \sim n \frac{\alpha_{\rm mom} b_{\rm EPJ}}{B_{\rm host}}.
  \label{eq:model_lt_mom}
\end{equation}
The coefficients for equation~(\ref{eq:model_lt_mom}) are summarized in
table~\ref{tab:model_lt_mom}.

\begin{table}
  \caption{Coefficient for equation~(\ref{eq:model_lt_mom})}
  \label{tab:model_lt_mom}
  \begin{tabular}{ll}
    \toprule
    $\alpha_{\rm mom}$  & 1.8 \\
  \bottomrule
  \end{tabular}
\end{table}

\subsubsection{Model of $T_{\rm const\, gt}$}

In the current implementation of FDPS, even if MPI is not used, the
global tree is constructed. The procedures of the construction of the
global tree is essentially the same as those of the local tree, except
for reordering particles. In reordering particles, EPJ and SPJ in all
LETs are first written to separate arrays. The indices of the arrays
for EPJ and SPJ are also saved in order to efficiently reorder the
particles at the reusing step. Thus $T_{\rm const\, gt}$ is given by
\begin{eqnarray}
  T_{\rm const\, gt} & \sim & 
  T_{\rm key\, gt} + T_{\rm sort\, gt} + T^{\rm const}_{\rm reorder\, gt} + T_{\rm link\, gt} \\
  T_{\rm key\, gt} & \sim & \frac{n_{\rm LET}  \alpha_{\rm key} \left(b_{\rm key}+b_{\rm EPJ} \right)}{B_{\rm host}} , \label{eq:gt_key} \\
  T_{\rm sort\, gt} & \sim  & \frac{\left(n+ n_{\rm LET} \right) \alpha_{\rm sort} b_{\rm key}}{B_{\rm host}} , \\
  T^{\rm const}_{\rm reorder\, gt} & \sim & \frac{\left(n+ n_{\rm LET}\right) \alpha^{\rm const}_{\rm reorder\, gt} \left( 2b_{\rm EPJ} + b_{\rm index} \right) }{B_{\rm host}} , \\ \label{eq:gt_sort_ptcl_const}
  T_{\rm link\, gt} & \sim &  \frac{ \frac{\left(n+ n_{\rm LET}\right)}{4} \alpha_{\rm link} \log_2\left( \frac{\left(n+ n_{\rm LET}\right)}{4} \right)}{B_{\rm host}} ,  \label{eq:gt_link}
\end{eqnarray}
where $n_{\rm LET}$ is the number of LETs and $b_{\rm index}$ is the
size of one index for EPJ and SPJ in bytes. Note that for the case of
center-of-mass approximation used here, the type of SPJ is the same as
that of EPJ and thus the size of SPJ is equal to $b_{\rm EPJ}$. The
coefficients for equations~(\ref{eq:gt_key}) to (\ref{eq:gt_link}) are
listed in table~\ref{tab:model_gt0}. We can see that $\alpha^{\rm
  const}_{\rm reorder\, gt}$ is larger than $\alpha_{\rm reorder\,
  lt}$ because for each node of LET we need to determine whether it is
EPJ or SPJ.

\begin{table}
  \caption{Coefficient for equations~(\ref{eq:gt_key}) to (\ref{eq:gt_link})}
  \label{tab:model_gt0}
  \begin{tabular}{ll}
    \toprule
    $\alpha_{\rm key}$ & 0.85 \\
    $\alpha^{\rm const}_{\rm reorder\, gt}$ & 5.0 \\
    $b_{\rm index}$ & 4 byte \\
  \bottomrule
  \end{tabular}
\end{table}

\subsubsection{Model of $T_{\rm reorder\, gt}$}

At the reusing step, we also reorder the particle in Morton order of
the global tree by using the index of arrays for EPJ and SPJ
constructed at the construction step. Thus $T_{\rm const\, gt}$ is
dominated by the times to read the indices, EPJ and SPJ once and that
to write EPJ and SPJ once and therefore $T_{\rm const\, gt}$ is given
by
\begin{equation}
  T^{\rm reuse}_{\rm reorder\, gt} \sim \frac{\left(n+ n_{\rm LET}\right) \alpha^{\rm reuse}_{\rm reorder\, gt} \left( 2b_{\rm EPJ}+b_{\rm index} \right)}{B_{\rm host}}. \\ \label{eq:gt_sort_ptcl_reuse}  
\end{equation}

The coefficients for equation~(\ref{eq:gt_sort_ptcl_reuse}) are listed
in table~\ref{tab:model_gt}. We can see that $\alpha^{\rm reuse}_{\rm
  reorder\, gt}$ is much smaller than $\alpha^{\rm const}_{\rm
  reorder\, gt}$ because we use the indices of the arrays of EPJ and
SPJ saved at the construction step to reorder the particles.

\begin{table}
  \caption{Coefficient for equation~(\ref{eq:gt_sort_ptcl_reuse})}
  \label{tab:model_gt}
  \begin{tabular}{ll}
    \toprule
    $\alpha^{\rm reuse}_{\rm reorder\, gt}$ & 1.0 \\
  \bottomrule
  \end{tabular}
\end{table}

\subsubsection{Model of $T_{\rm mom\, gt}$}

The procedure of the calculation of the multipole moments of the cells
of the global tree is almost the same as that of the local tree. Thus
$T_{\rm mom\, gt}$ is given by
\begin{eqnarray}
  T_{\rm mom\, gt} \sim \frac{\left( n + n_{\rm LET} \right) \alpha_{\rm mom} b_{\rm EPJ}}{B_{\rm host}}.
\end{eqnarray}

\subsubsection{Models of $T_{\rm force, const}$ and $T_{\rm force, reuse}$}
The elapsed times for the force calculation at the construction step
$T_{\rm force, const}$ and at the reusing step $T_{\rm force, reuse}$
are given by
\begin{eqnarray}
  T_{\rm force, const} &\sim& \max \left( T_{\rm cp\, all},\: T_{\rm H2D\, all} \right) \nonumber \\
    && + \max \left( T_{\rm kernel},\: T_{\rm H2D\, EPI}+T_{\rm H2D\, list},\: T_{\rm D2H\, FORCE},\: T_{\rm const\, list} + T_{\rm cp\, EPI}+T_{\rm cp\, list}+T_{\rm wb+int+cp}  \right)  \nonumber \\
    && + T_{\rm cp\, FORCE}, \label{eq:t_force_const} \\
  T_{\rm force, reuse} &\sim& \max \left( T_{\rm cp\, all},\: T_{\rm H2D\, all} \right) \nonumber \\
  && + \max \left( T_{\rm kernel},\: T_{\rm H2D\, EPI},\: T_{\rm D2H\, FORCE},\: T_{\rm cp\, EPI}+T_{\rm wb+int+cp} \right) \nonumber \\
  && + T_{\rm cp\, FORCE}, \label{eq:t_force_reuse}
\end{eqnarray}
where $T_{\rm cp\, all}$, $T_{\rm H2D\, all}$, $T_{\rm kernel}$,
$T_{\rm H2D\, EPI}$, $T_{\rm H2D\, list}$, $T_{\rm D2H\, FORCE}$,
$T_{\rm const\, list}$, $T_{\rm cp\, EPI}$, $T_{\rm cp\, list}$ and
$T_{\rm wb+int+cp}$ are the times for copying of EPJs and SPJs to the
send buffer, sending EPJs and SPJs from the host to GPGPU, the force
kernel on GPGPU, sending EPIs to GPGPU, sending the interaction lists
to GPGPU, receiving FORCEs from GPGPU, constructing the interaction
list, copying EPIs to the send buffer, copying the interaction lists,
and copying the data of FORCEs to FPs, integrating orbits of FPs and
copying the data of FPs to EPIs and EPJs, respectively.

Each components in equations~(\ref{eq:t_force_const}) and (\ref{eq:t_force_reuse}) are given by
\begin{eqnarray}
  T_{\rm cp\, all}  & \sim &  \frac{\left( n + n_{\rm LET} \right) \alpha_{\rm cp\, all} \left( b_{\rm EPJ}+b_{\rm EPJ\, buf} \right)}{B_{\rm host}} , \label{eq:time_cp_all} \\ 
  T_{\rm H2D\, all}  & \sim &  \frac{\left (n + n_{\rm LET} \right) \alpha_{\rm H2D\, all} b_{\rm EPJ\, buf} }{B_{\rm transfer}} , \\
  T_{\rm kernel} & \sim & n \langle n_{\rm list} \rangle \left( \frac{\alpha_{\rm GPU, kernel} n_{\rm op}}{F_{\rm GPU}}
  + \frac{\alpha_{\rm GPU, mm} \left( b_{\rm ID} + b_{\rm EPJ} \right) }{\langle n_i \rangle B_{\rm GPU}} \right),  \label{eq:kernel} \\  
  T_{\rm H2D\, EPI}  & \sim &  \frac{n \alpha_{\rm H2D\, EPI} b_{\rm EPI\, buf}}{B_{\rm transfer}}, \\
  T_{\rm H2D\, list} & \sim &  \frac{ n \langle n_{\rm list} \rangle \alpha_{\rm H2D\, list} b_{\rm ID}}{\langle n_i \rangle B_{\rm transfer}} , \\
  T_{\rm D2H\, FORCE} & \sim & \frac{n \alpha_{\rm D2H, FORCE} b_{\rm FORCE\, buf}}{B_{\rm transfer}} ,  \\
  T_{\rm const\, list}  & \sim &
  \frac{n \langle n_{\rm list} \rangle \alpha_{\rm const\, list} b_{\rm ID} }{\langle n_{\rm i} \rangle B_{\rm host}}, \label{eq:time_list} \\
  T_{\rm cp\, EPI}   & \sim &  \frac{n \alpha_{\rm cp\, EPI} \left( b_{\rm EPI}+b_{\rm EPI\, buf} \right) }{B_{\rm host}} , \\
  T_{\rm cp\, list}  & \sim &  \frac{n \langle n_{\rm list} \rangle  \alpha_{\rm cp\, list} b_{\rm ID}}{\langle n_i \rangle B_{\rm host}}, \\ 
  T_{\rm wb+int+cp} & \sim & \frac{n \alpha_{\rm wb+int+cp} \left( b_{\rm FORCE}+b_{\rm FP}+b_{\rm EPI}+b_{\rm EPJ} \right) }{B_{\rm host}}, \label{eq:wb_int_cp} \\
  T_{\rm cp\, FORCE} & \sim  & \frac{n \alpha_{\rm cp\, FORCE} \left( b_{\rm FORCE}+b_{\rm FORCE\, buf} \right)}{B_{\rm host}} , \\
  \langle n_{\rm list} \rangle & \sim & \langle n_{\rm i} \rangle + \frac{14\langle n_{\rm i} \rangle^{2/3}}{\theta} + \frac{21\pi \langle n_{\rm i} \rangle^{1/3}}{\theta^2} + \frac{28\pi}{3\theta^3} {\rm log_2} \left[ \frac{\theta}{2.8}\left(n^{1/3} - \langle n_{\rm i} \rangle^{1/3}\right)\right], \label{eq:n_list}  
\end{eqnarray}
where $b_{\rm EPI(J)\, buf}$, $b_{\rm ID}$, $b_{\rm FORCE}$ and
$b_{\rm FORCE\, buf}$ are the data size of EPI(J) in the send buffer,
index of EPJ and SPJ in the interaction list, FORCE and FORCE in the
receive buffer, $B_{\rm transfer}$ and $B_{\rm GPU}$ are the effective
bandwidths of the data bus between the host and GPGPU and the main
memory of GPGPU, $F_{\rm GPU}$ is the peak speed of floating point
operation of GPGPU, $n_{\rm op}$ is the number of floating point
operation per interaction and $\langle n_{\rm list} \rangle$ is the
average size of the interaction list. The elapsed times are determined
by the bandwidth of the main memory of the host or the data bus
between the host and the GPGPU, except for the time for the force
calculation, $T_{\rm kernel}$. The model of $T_{\rm kernel}$ is a bit
complicated. We will describe how to construct this model in
appendix~\ref{sec:model_gpu}.

We summarize the coefficients for equations~(\ref{eq:time_cp_all}) to
(\ref{eq:wb_int_cp}) in table \ref{tab:time_force}. We can see that
$\alpha_{\rm list}$ and $\alpha_{\rm GPU\, transfer}$ are much larger
than unity because the random access of the main memory is slow both
the host and the GPGPU.

\begin{table}[h]
  \caption{Coefficients for equations~(\ref{eq:time_cp_all}) to
    (\ref{eq:wb_int_cp})}
  \label{tab:time_force}
  \begin{tabular}{ll}
    \toprule
    $\alpha_{\rm cp\, all}$    & 0.87 \\ 
    $\alpha_{\rm H2D\, all}$   & 1.0  \\ 
    $\alpha_{\rm cp\, EPI}$    & 1.0  \\ 
    $\alpha_{\rm H2D\, EPI}$   & 1.0  \\
    $\alpha_{\rm cp\, list}$   & 1.0  \\ 
    $\alpha_{\rm H2D\, list}$  & 0.86 \\ 
    $\alpha_{\rm cp\, FORCE}$  & 1.2  \\ 
    $\alpha_{\rm D2H\, FORCE}$  & 1.2 \\ 
    $\alpha_{\rm GPU\, calc}$  & 1.7  \\ 
    $\alpha_{\rm GPU\, transfer}$    & 22  \\ 
    $\alpha_{\rm const\, list}$ & 19  \\
    $\alpha_{\rm wb+int+cp}$   & 1.4  \\
    $b_{\rm EPJ\, buf}$ & 16 byte \\
    $b_{\rm EPI\, buf}$ & 12 byte \\
    $b_{\rm ID}$ & 4 byte \\
    $b_{\rm ID\, buf}$ & 4 byte \\
    $b_{\rm FORCE}$   & 32 byte \\
    $b_{\rm FORCE\, buf}$ & 16 byte \\
    $n_{\rm op}$ & 23 \\
  \bottomrule
  \end{tabular}
\end{table}

\subsubsection{Model of $T_{\rm step, const}$ and $T_{\rm step, reuse}$}

In the previous sections, we made the performance models of each step
of the interaction calculation. By using these models, the total
execution time is expressed by equations (\ref{eq:total_const}) for
the construction step and by equation (\ref{eq:total_reuse}) for the
reusing step.

\begin{eqnarray}
  T_{\rm step, const} &\sim&  \bigg[ n ( \alpha_{\rm root} b_{\rm FP}
    + \alpha_{\rm key} \left( b_{\rm FP} + b_{\rm key} \right) 
    + \alpha_{\rm reorder} \left( 2b_{\rm FP} + b_{\rm EPI} + b_{\rm EPJ} + b_{\rm key} \right) \nonumber \\
    && + \alpha_{\rm cp\, FORCE} \left( b_{\rm FORCE} + b_{\rm FORCE\, buf} \right)  ) \nonumber  \\
    && + (n+n_{\rm LET}) \left( \alpha_{\rm key} \left( b_{\rm EPJ} + b_{\rm key} \right)
    + \alpha^{\rm const}_{\rm reorder\, gt} \left( 2b_{\rm EPJ} + b_{\rm index} \right) \right) \nonumber  \\
    && + (2n+n_{\rm LET})(\alpha_{\rm sort} 24 b_{\rm key} + \alpha_{\rm mom} b_{\rm EPJ} ) \bigg] / B_{\rm host} \nonumber \\
  && + \alpha_{\rm link} \left( \left( n/4 \right) \log_2\left( n/4\right) + \left( \left(2n+n_{\rm LET}\right)/4 \right) \log_2\left( \left(2n+n_{\rm LET}\right)/4\right) \right) / B_{\rm host} \nonumber \\
  && + (n+n_{\rm LET}) \max \bigg[ \alpha_{\rm cp\, all} \left( b_{\rm EPJ} + b_{\rm EPJ\, buf}\right) / B_{\rm host}, \alpha_{\rm H2D\, all} b_{\rm EPJ\, buf} / B_{\rm transfer} \bigg] \nonumber  \\
  && + n \max \Bigg[ n_{\rm list} \left( \frac{\alpha_{\rm GPU\, calc} n_{\rm op}}{F_{\rm GPU}} + \frac{\alpha_{\rm GPU} \left( b_{\rm ID} + b_{\rm EPJ} \right) }{\langle n_i \rangle B_{\rm GPU}}\right), \nonumber \\
    && \Big (\alpha_{\rm H2D\, EPI} b_{\rm EPI\, buf}
    +  \frac{n_{\rm list}}{\langle n_{\rm i}\rangle} \alpha_{\rm H2D\, list} b_{\rm ID\, buf} \Big) / B_{\rm transfer}, \nonumber  \\
    && \alpha_{\rm D2H\, FORCE} b_{\rm FORCE\, buf} / B_{\rm transfer}, \nonumber \\
    && \Big( \alpha_{\rm cp\, EPI} \left(b_{\rm EPI} + b_{\rm EPI\, buf} \right)
    +  \frac{n_{\rm list}}{\langle n_{\rm i}\rangle} 
    \big( \alpha_{\rm cp\, list} b_{\rm ID} + \alpha_{\rm const\, list} b_{\rm ID} \big) \nonumber \\ 
    && + \alpha_{\rm wb+int+cp} \left( b_{\rm FORCE} + b_{\rm FP} + b_{\rm EPI} + b_{EPJ} \right)   \Big) / B_{\rm host} \Bigg].
  \label{eq:total_const}
\end{eqnarray}

\begin{eqnarray}
  T_{\rm step, reuse} &\sim&  \bigg[ n \alpha_{\rm cp\, FORCE} \left( b_{\rm FORCE} + b_{\rm FORCE\, buf} \right)
    + (n+n_{\rm LET}) \alpha^{\rm reuse}_{\rm reorder\, gt} \left( 2b_{\rm EPJ} + b_{\rm index} \right) \nonumber \\
    && + (2n+n_{\rm LET}) \alpha_{\rm mom} b_{\rm EPJ} \bigg] / B_{\rm host} \nonumber \\
  && + (n+n_{\rm LET}) \max \bigg[ \alpha_{\rm cp\, all} \left( b_{\rm EPJ} + b_{\rm EPJ\, buf} \right)  / B_{\rm host}, \alpha_{\rm H2D\, all} b_{\rm EPJ\, buf} / B_{\rm transfer} \bigg] \nonumber  \\
  && + n \max \Bigg[ n_{\rm list} \left( \frac{\alpha_{\rm GPU\, calc} n_{\rm op}}{F_{\rm GPU}} + \frac{\alpha_{\rm GPU\, transfer} \left(b_{\rm ID} + b_{\rm EPJ} \right) }{\langle n_i \rangle B_{\rm GPU}} \right), \nonumber \\
    && \alpha_{\rm H2D\, EPI} b_{\rm EPI\, buf} / B_{\rm transfer}, \nonumber  \\
    && \alpha_{\rm D2H\, FORCE} b_{\rm FORCE\, buf} / B_{\rm transfer}, \nonumber \\
    && \Big( \alpha_{\rm cp\, EPI} \left( b_{\rm EPI} + b_{\rm EPI\, buf}\right)
    + \alpha_{\rm wb+int+cp} \left( b_{\rm FORCE} + b_{\rm FP} + b_{\rm EPI} + b_{\rm EPJ} \right)   \Big) / B_{\rm host} \Bigg].  
 \label{eq:total_reuse}
\end{eqnarray}

Thus, the mean execution time per step $T_{\rm step, single}$,
substituting the efficiency parameters $\alpha$s and the size
parameters $b$s with the measured values listed in tables is given by
\begin{eqnarray}
  T_{\rm step, single} & \sim & \frac{1}{n_{\rm reuse}}\left( T_{\rm step, const} + (n_{\rm reuse}-1)T_{\rm step, reuse} \right) \nonumber \\
  &\sim & 68\left(n+n_{\rm LET} \right)/B_{\rm host} + \left( 174n + 58n_{\rm LET} \right) \max \left( 42/B_{\rm host}, 16/B_{\rm transfer} \right) \nonumber \\
  && + n \max \Big( n_{\rm list} \left( 39.1/F_{\rm GPU} + 440/(\langle n_i \rangle B_{\rm GPU}) \right), 19.2/B_{\rm transfer}, 282 / B_{\rm host} \Big) \nonumber \\
  && + \frac{1}{n_{\rm reuse}} \Bigg[ \Big( 1580n + 1157n_{\rm LET} \nonumber \\
  &&  + 54.4 \left( \left( n/4 \right) \log_2\left( n/4\right) + \left( \left(2n+n_{\rm LET}\right)/4 \right) \log_2\left( \left(2n+n_{\rm LET}\right)/4\right) \right) \Big) / B_{\rm host} \nonumber \\
    &&  + n \Bigg( \max \Big( n_{\rm list} \left( 39.1/F_{\rm GPU} + 440/(\langle n_i \rangle B_{\rm GPU}) \right), 
    \left( 12 + 3.4 n_{\rm list}/\langle n_i \rangle  \right) / B_{\rm transfer}, \nonumber \\
    && 19.2/B_{\rm transfer}, \left( 282 + 80n_{\rm list}/\langle n_i \rangle \right) / B_{\rm host} \Big), \nonumber \\
    && - \max \Big( n_{\rm list} \left( 39.1/F_{\rm GPU} + 440/(\langle n_i \rangle B_{\rm GPU}) \right),
    19.2/B_{\rm transfer}, 282 / B_{\rm host} \Big) \Bigg) \Bigg]
    \label{eq:total_single}
\end{eqnarray}

\begin{figure}
  \begin{center}
    \includegraphics[width=14cm]{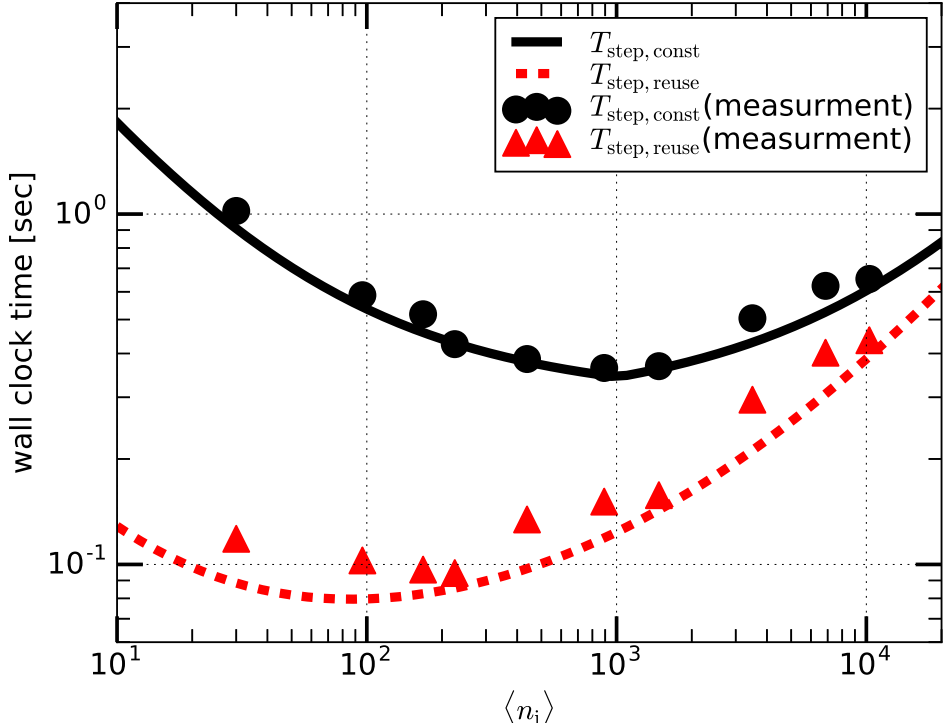}
  \end{center}
  \caption{Total elapsed time in the cases of both the construction
    and reusing steps. Solid (dashed) curve and Filled circles
    (triangles) indicate respectively the predicted time by using the
    model and the measured times, in the case of the construction
    (reusing) step.}
  \label{fig:compare}
\end{figure}

In order to check whether the model we constructed is reasonable or
not, we compared the time predicted by equations
(\ref{eq:total_const}) and (\ref{eq:total_reuse}) with the measured
times in figure \ref{fig:compare}. The predicted times agree with the
measured times very well.

In the following, we analyze the performance of the $N$-body
simulations on hypothetical systems with various ${B_{\rm host}}$,
${B_{\rm transfer}}$, ${F_{\rm GPU}}$ and ${B_{\rm GPU}}$ by using the
performance model. Unless otherwise noted, we assume the hypothetical
system with ${B_{\rm host}}=100$ GB/s, ${B_{\rm transfer}}=10$ GB/s,
${B_{\rm GPU}}=500$ GB/s and ${ F_{\rm GPU}}=10$ Tflops as a reference
system. This reference system can be regarded as a modern HPC system
with a high-end accelerator.

\begin{figure}
  \begin{center}
    \includegraphics[width=16cm]{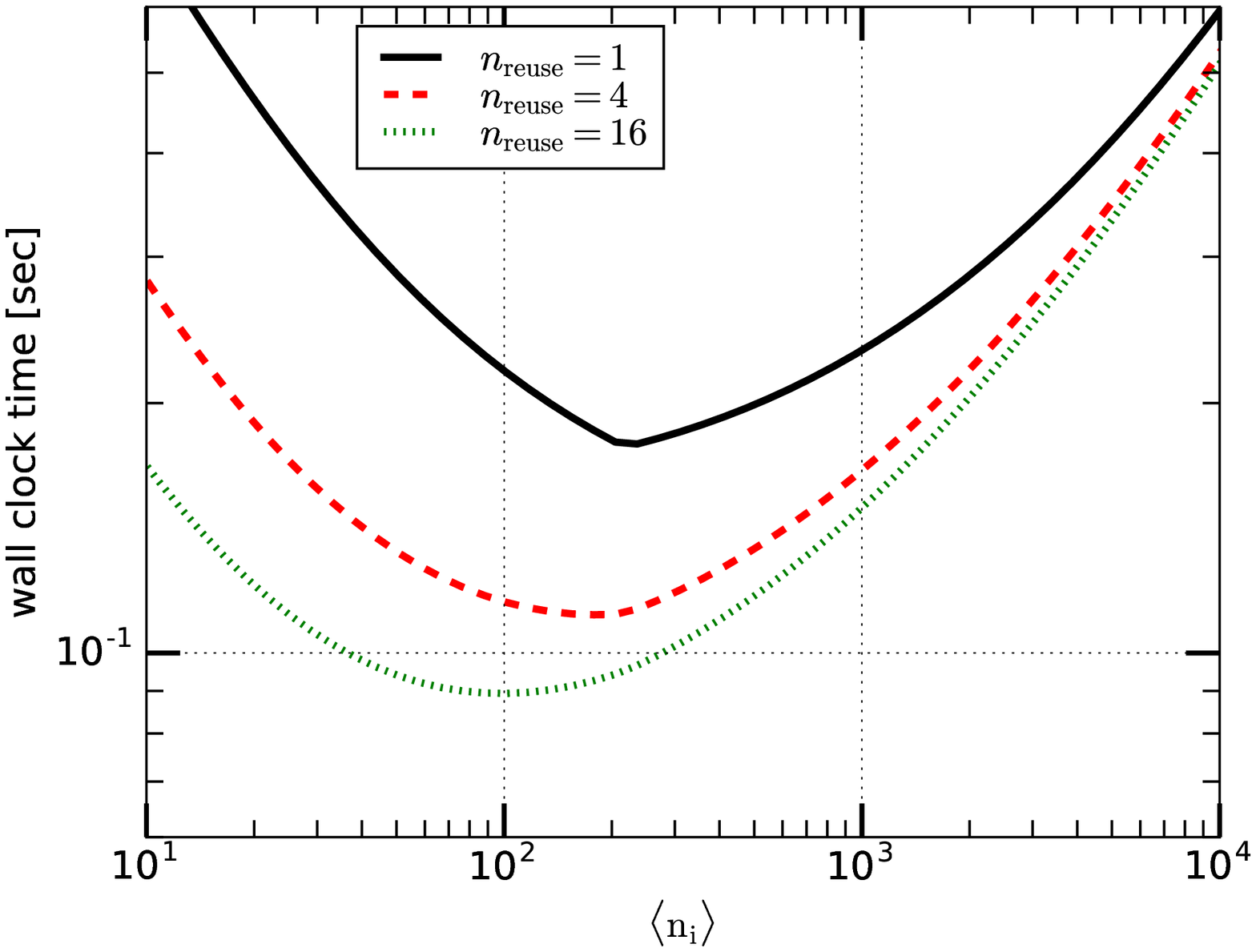}
  \end{center}
  \caption{Calculation time per timestep on the reference system for
    various $n_{\rm reuse}$.}
  \label{fig:compare_total_n_reuse}
\end{figure}

Figure~\ref{fig:compare_total_n_reuse} shows the calculation time per
timestep on the reference system for 4M particles and $\theta=0.5$ for
$n_{\rm reuse}=$ 1, 4 and 16. We can see that the difference in the
performance for $n_{\rm reuse}=4$ and $n_{\rm reuse}=16$ is relatively
small. In the rest of the section, we use $n_{\rm reuse}=16$ and
$\theta=0.5$.

\begin{figure}
  \begin{center}
    \includegraphics[width=16cm]{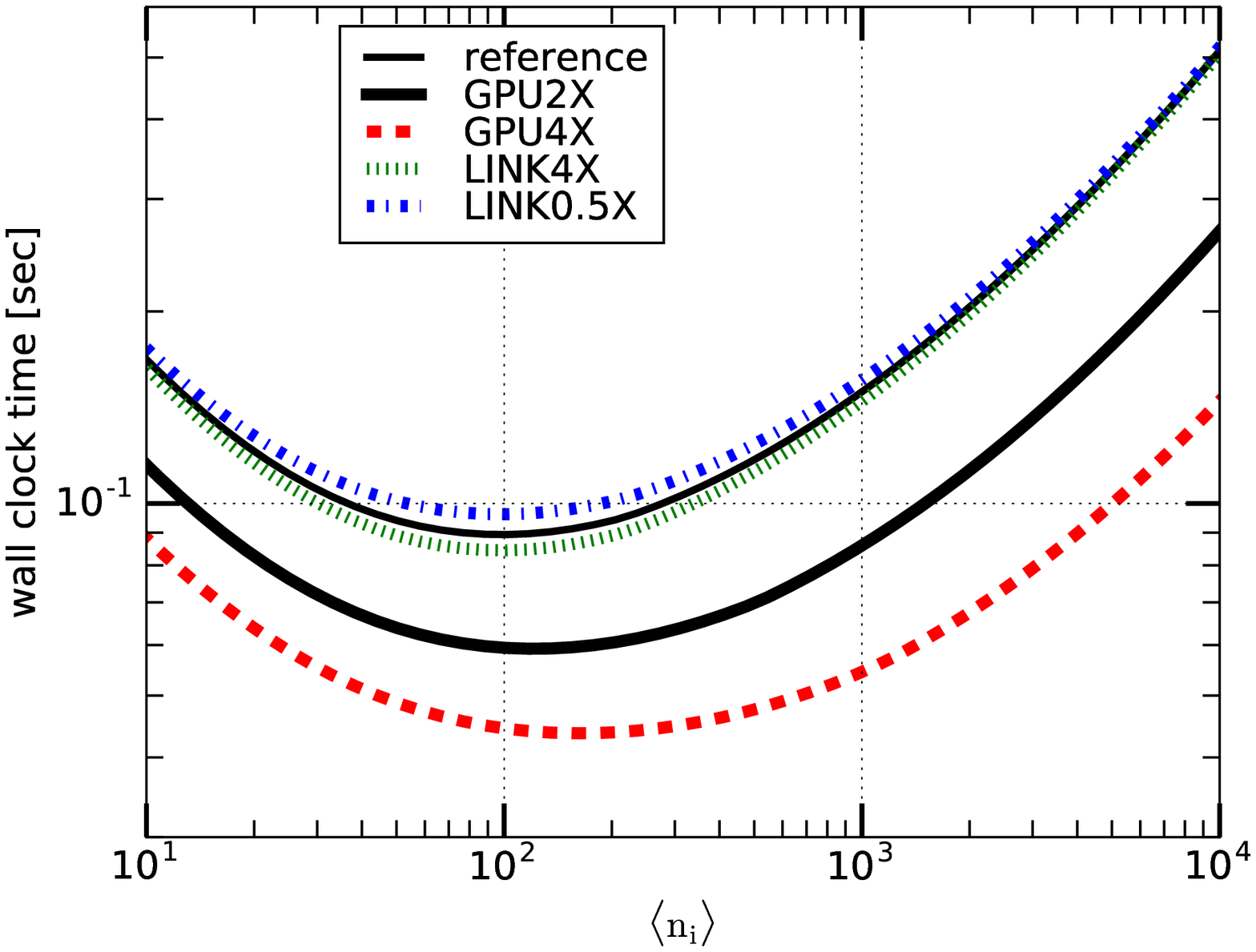}
  \end{center}
  \caption{The mean execution time per step, $T_{\rm step\, single}$,
    on various hypothetical systems against $\langle n_{\rm i}
    \rangle$.}
  \label{fig:compare_total_hyp}
\end{figure}

We consider four different types of hypothetical systems: GPU2X,
GPU4X, LINK4X and LINK0.5X. Their parameters are listed in
table~\ref{tab:all_system}. Figure~\ref{fig:compare_total_hyp} shows
the calculation times per timestep for the four hypothetical systems.
We can see that increasing the bandwidth between CPU and accelerator
(LINK4X) has relatively small effect on the performance. On the other
hand, increasing overall accelerator performance has fairly large
impact.

\begin{table}
  \begin{tabular}{ccccc}
    \toprule system & ${F_{\rm GPU}}$ & ${B_{\rm host}}$ & ${B_{\rm GPU}}$ & ${B_{\rm transfer}}$ \\
    \midrule
    reference & 10 Tflops & 100 GB/s & 500 GB/s & 10 GB/s \\
    GPU2X     & 20 Tflops & 100 GB/s & 1   TB/s & 10 GB/s \\
    GPU4X     & 40 Tflops & 100 GB/s & 2   TB/s & 10 GB/s \\
    LINK4X    & 10 Tflops & 100 GB/s & 500 GB/s & 40 GB/s \\
    LINK0.5X    & 10 Tflops & 100 GB/s & 500 GB/s & 5  GB/s \\
  \bottomrule
  \end{tabular}
  \caption{Parameters of hypothetical systems.}
  \label{tab:all_system}
\end{table}

\begin{figure}
  \includegraphics[width=1.0\hsize]{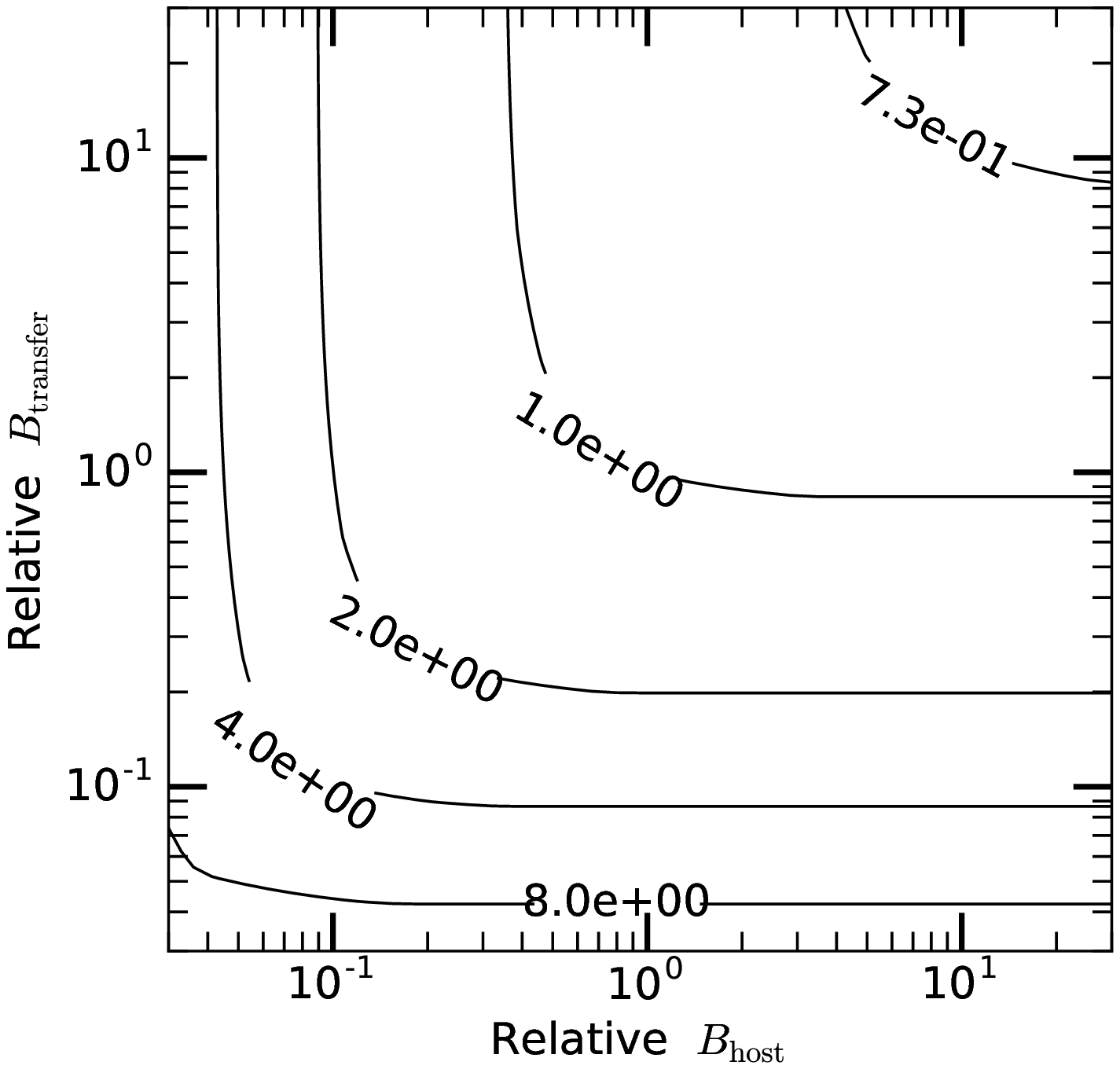}
  \caption{Relative execution time of hypothetical systems in the
    two-dimensional plane of relative $B_{\rm host}$ and relative
    $B_{\rm transfer}$. The values of contours, x- and y- axes are
    normalized by those of the reference system.}
  \label{fig:2d_map}
\end{figure}

Figure~\ref{fig:2d_map} shows the relative execution time of
hypothetical systems in the two-dimensional plane of $B_{\rm host}$
and $B_{\rm transfer}$.  We can see that the increase of $B_{\rm
  host}$ or $B_{\rm transfer}$, or even both, would not give
significant performance improvement, while the increase of the
accelerator performance gives significant speedup. Even when both
$B_{\rm host}$ and $B_{\rm transfer}$ are reduced by a factor of 10,
overall speed is reduced by a factor of 4. Thus, if the speed of
accelerator is improved by a factor of 10, the overall speedup is 4.
Thus we can conclude that the current implementation of FDPS can
provide good performance not only on the current generation of GPGPUs
but also future generations of GPGPUs or other accelerator-based
systems, which will have relatively poor data transfer bandwidth
compare to the calculation performance.

\subsection{Performance Model on Multiple Nodes}
\label{sec:perf_model_para}

In this section, we discuss the performance model of the parallelized
$N$-body simulation with method G2. Here, we assume the network is the
same as that of K computer. Detailed communication algorithms are
given in \citet{Iwasawaetal2016}.

The time per step is given by
\begin{equation}
  T_{\rm step, para} = T_{\rm step, single} + T_{\rm exch}/n_{\rm reuse} + T_{\rm dc}/n_{\rm dc} + T_{\rm LET, const}/n_{\rm reuse} + T_{\rm LET, exch},
\end{equation}
where $T_{\rm exch}$, $T_{\rm dc}$ $T_{\rm LET, const}$ and $T_{\rm
  LET, exch}$ are the times for the exchange of particles, the domain
decomposition, the construction of LETs and the exchange of LETs and
$n_{\rm dc}$ is the number of steps for which the same domain
decomposition is used. We consider the case when particles do not move
much in a single step and thus ignore $T_{\rm exch}$.

The time for the domain decomposition is given by
\begin{equation}
  T_{\rm dc} \sim T_{\rm dc,\, collect} + T_{\rm dc,\, sort},
\end{equation}
where $T_{\rm dc, collect}$ and $T_{\rm dc, sort}$ are the times to
collect sample particles to root processes and to sort particles on
the root processes.

According to \citet{Iwasawaetal2016}, $T_{\rm dc,\, collect}$ and
$T_{\rm dc,\, sort}$ are given by
\begin{eqnarray}
    T_{\rm dc,\, collect} &\sim& n^{1/6}_{\rm p} \tau_{\rm gather, startup} + \frac{n_{\rm smp} n^{2/3}_{\rm p} \alpha_{\rm gather} b_{\rm pos}}{B_{\rm inj}}, \label{eq:dc_collect}  \\
  T_{\rm dc,\, sort} &\sim& \frac{n_{\rm smp} n^{2/3}_{\rm p} \log \left( n^3_{\rm smp} n^{5/3}_{\rm p} \right) \alpha_{\rm dc, sort} b_{\rm pos}}{B_{\rm host}}, \label{eq:dc_sort}
\end{eqnarray}
where $n_{\rm p}$ is the number of processes, $n_{\rm smp}$ is the
number of sample particles to determine the domains, $b_{\rm pos}$ is
the data size of the position of the particle and $B_{\rm inj}$ is the
effective injection bandwidth, $\tau_{\rm gather, startup}$ is the
startup time for {\tt MPI\_Gather} and $\alpha_{\rm gather}$ is the
efficiency parameter of communicating data with {\tt MPI\_Gather}. We
will describe how to measure the parameters $\tau_{\rm gather,
  startup}$ and $\alpha_{\rm gather}$ in
appendix~\ref{sec:allgather}. The coefficients for
equation~(\ref{eq:dc_sort}) are listed in table~\ref{tab:dc}. Since we
used the quick sort here, $\alpha_{\rm dc\, sort}$ is much larger than
unity.

\begin{table}[h]
  \caption{Coefficients for equations~(\ref{eq:dc_collect}) and (\ref{eq:dc_sort})}
  \label{tab:dc}
  \begin{tabular}{ll}
    \toprule
    $\alpha_{\rm dc\, sort}$ & 7.7 \\
    $\alpha_{\rm gather}$  & 0.62 \\
    $\tau_{\rm gather, startup}$  & $ 1.2 \times 10^{-5}$ sec  \\
    $b_{\rm pos}$  & 24 byte \\
  \bottomrule
  \end{tabular}
\end{table}

In the original implementation of FDPS, {\tt MPI\_Alltoallv} was used
for the exchange of LETs and it was the main bottleneck of the
performance for large $n_{\rm p}$ \citep{Iwasawaetal2016}. Thus,
recently, we developed a new algorithms for the exchange of LETs to
avoid the use of {\tt MPI\_Alltoallv} \citep{Iwasawaetal2018}. The new
algorithm is as follows:
\begin{enumerate}
\item Each process sends the multipole moment of the top level cell of
  its local tree to all processes using {\tt MPI\_Allgather}.
\item Each process calculates the distances from all other domains.
\item If the distance between process $i$ and $j$ is large enough so
  that process $i$ can be regarded as one cell from process $j$, that
  domain already has the necessary information. If not, process $i$
  construct LET for process $j$ and send it to process $j$. 
\end{enumerate}

With this new algorithm, the times for the exchange LET is expressed
as
\begin{eqnarray}
  T_{\rm LET, const} &\sim& \frac{\left( n_{\rm LET} - n_{\rm p} + n_{\rm p, close} \right) \alpha_{\rm const\, list} b_{\rm ID}}{B_{\rm host}}, \\
  T_{\rm LET, exch} &\sim& T_{\rm LET, allgather} + T_{\rm LET, p2p}, \\
  T_{\rm LET, allgather} &\sim& n^{1/4}_{\rm p} \tau_{\rm allgather, startup} + \frac{n_{\rm p} \alpha_{\rm allgather} b_{\rm EPJ}}{B_{\rm inj}} \label{eq:let_allgather}, \\
  T_{\rm LET, p2p} &\sim&  n_{\rm p, close} \tau_{\rm p2p, startup}  + \frac{2 \left( n_{\rm LET} - n_{\rm p} + n_{\rm p, close} \right) \alpha_{\rm p2p} b_{\rm EPJ}}{B_{\rm inj}}  \label{eq:let_p2p}, \\
  n_{\rm LET} &\sim& \frac{14n^{2/3}}{\theta} + \frac{21\pi n^{1/3}}{\theta^2} + \frac{28\pi}{3\theta^3} {\rm log_2} \left\{ \frac{\theta}{2.8}\left[\left(nn_{\rm p}\right)^{1/3} - n^{1/3}\right] \right\} + n_{\rm p} - n_{\rm p, close}, \\
  n_{\rm p, close} &\sim&  6\left( \frac{1}{\theta} + 1 \right) + 3\pi \left( \frac{1}{\theta} + 1 \right)^2 + \frac{4\pi}{3} \left( \frac{1}{\theta} + 1 \right)^3,
\end{eqnarray}
where $T_{\rm LET, allgather}$ and $T_{\rm LET, p2p}$ are the time for
the time for exchange LETs using {\tt MPI\_Allgather} and {\tt
  MPI\_Isend/Irecv}, $n_{\rm p, close}$ is the number of the processes
to exchange LETs with {\tt MPI\_Isend/Irecv}, $\tau_{\rm
  p2p,allgather}$ and $\tau_{\rm p2p,startup}$ are the startup times
for {\tt MPI\_Allgather} and {\tt MPI\_Isend/Irecv} and $\alpha_{\rm
  allgather}$ and $\alpha_{\rm p2p}$ are the efficiency parameters for
exchange LETs with {\tt MPI\_Allgather} and {\tt MPI\_Isend/Irecv},
respectively.

The parameters for equations~(\ref{eq:let_allgather}) and
(\ref{eq:let_p2p}) are listed in table~\ref{tab:coef2}.

\begin{table}[h]
  \caption{Coefficients for equations~(\ref{eq:let_allgather}) and (\ref{eq:let_p2p})}
  \label{tab:coef2}
  \begin{tabular}{ll}
    \toprule
    $\alpha_{\rm allgather, ptcl}$  & 0.62 \\
    $\alpha_{\rm p2p}$  & 1.0 \\
    $\tau_{\rm allgather, startup}$  & $ 1.2 \times 10^{-5}$ sec \\
    $\tau_{\rm p2p, startup}$ & $1.0 \times 10^{-5}$ sec  \\
  \bottomrule
  \end{tabular}
\end{table}

\begin{figure}
    \begin{center}
      \includegraphics[width=18cm]{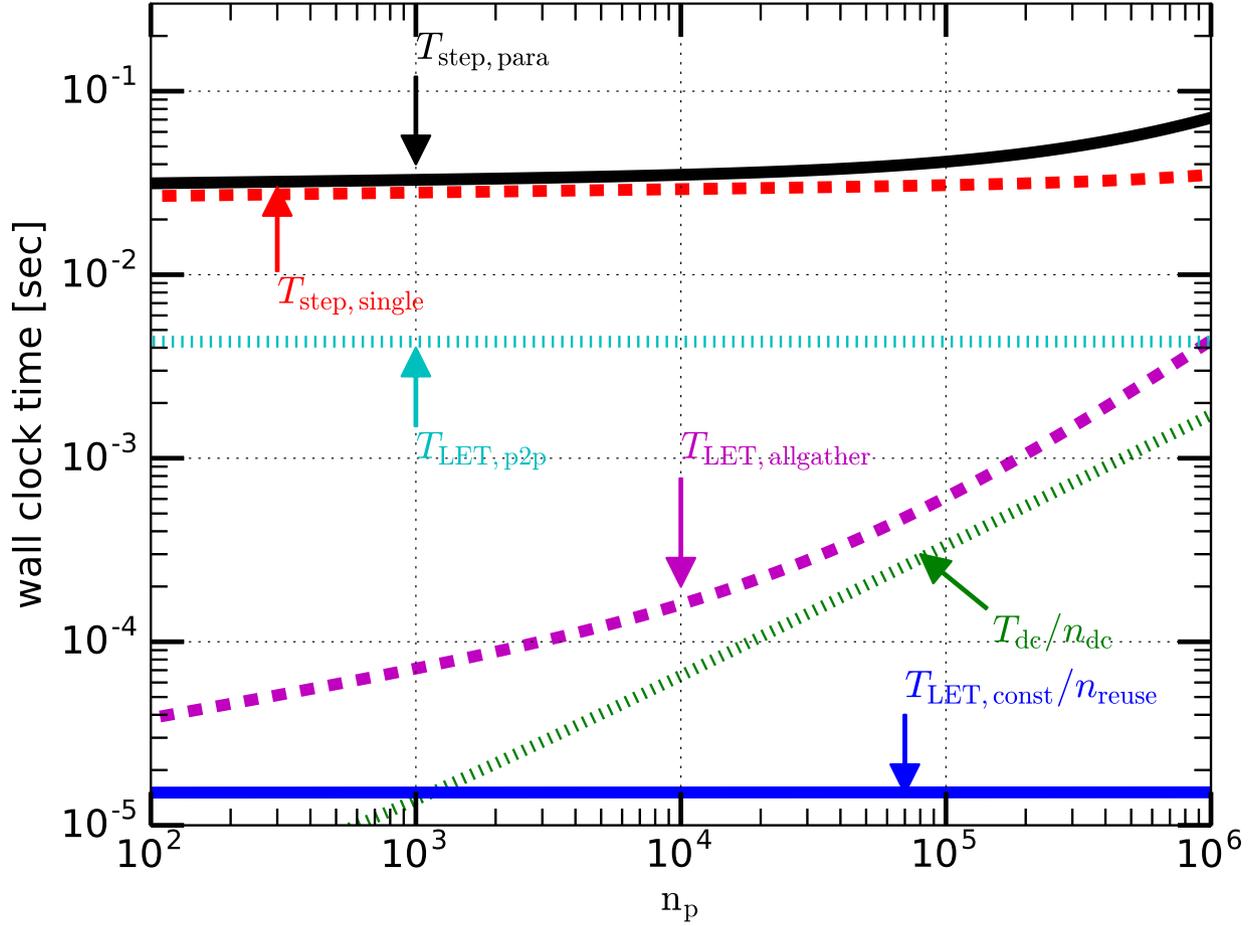}
    \end{center}
    \caption{Weak-scaling calculation time plotted as a function of
      $n_{\rm p}$. The number of particles per node is $2^{20}$.}
  \label{fig:total_parallel_break_down_weak}
\end{figure}

Figure~\ref{fig:total_parallel_break_down_weak} shows the weak scaling
performance for $N$-body simulations in the case of the number of
particles per process of $n=2^{20}$. Here we assume that $n_{\rm
  reuse}=n_{\rm dc}=16$, $n_{\rm smp}=30$, $\langle n_{\rm i}
\rangle=250$ and $\theta=0.5$. We can see that $T_{\rm step, para}$ is
almost constant for $n_{\rm p} \lesssim 10^5$. For $n_{\rm p} \gtrsim
10^5$, $T_{\rm step, para}$ slightly increases because $T_{\rm LET,
  allgather}$ becomes large. Roughly speaking, when $n$ is much larger
than $n_{\rm p}$, the parallel efficiency is high.

\begin{figure}
  \begin{center}
    \begin{tabular}{c}
      
      \begin{minipage}{0.5\hsize}
        \begin{center}
          \includegraphics[width=9cm]{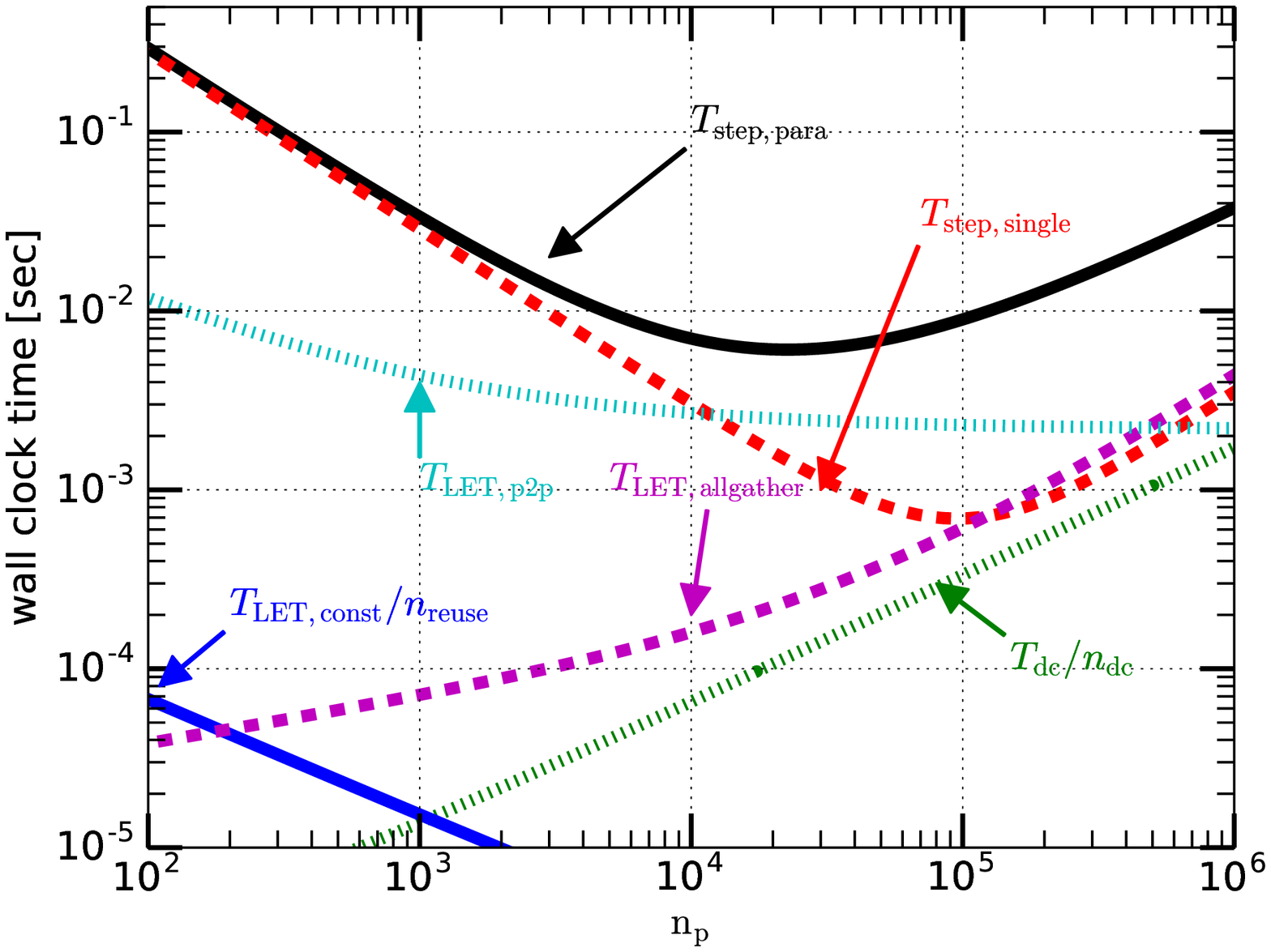}
        \end{center}
      \end{minipage}

      \begin{minipage}{0.5\hsize}
        \begin{center}
          \includegraphics[width=9cm]{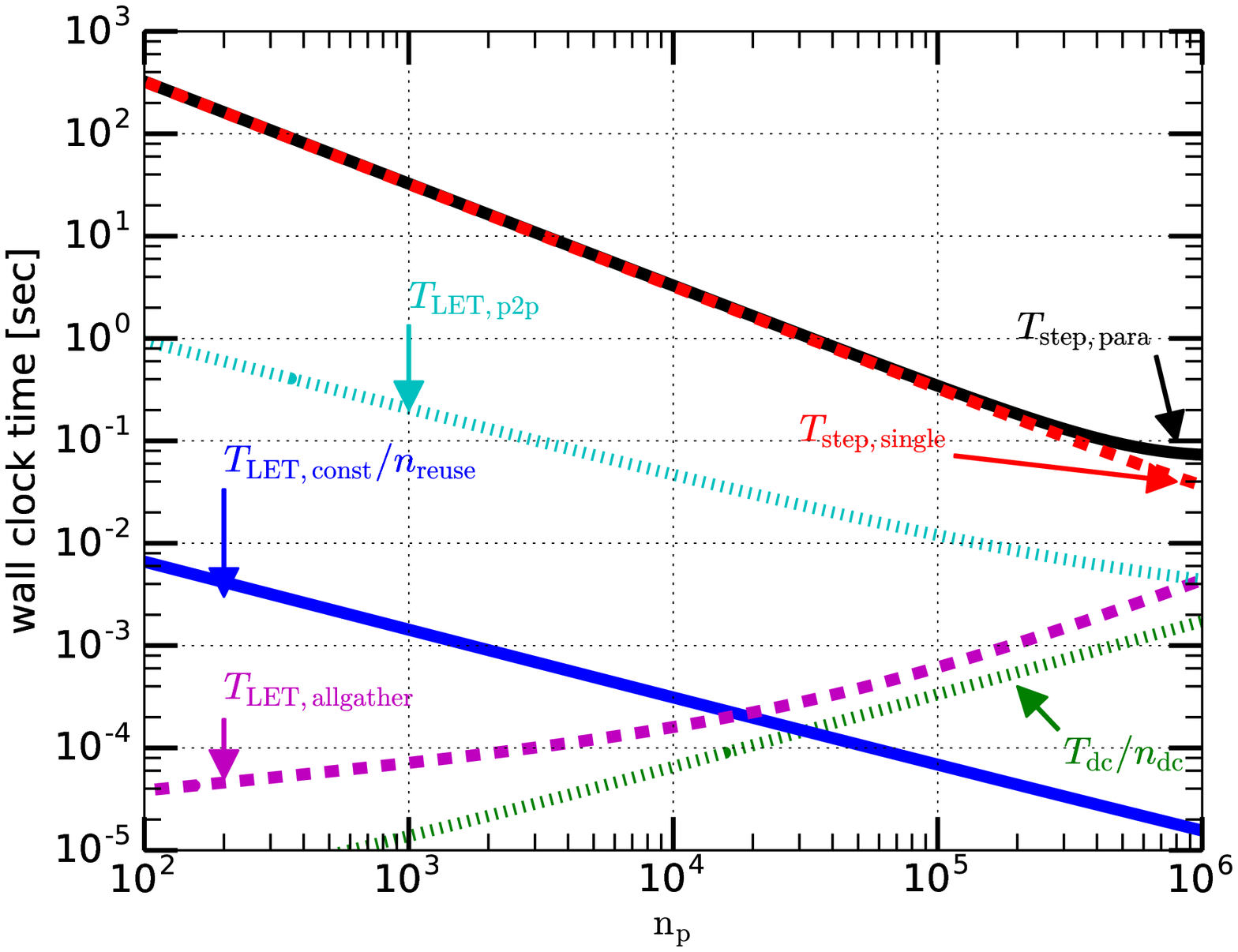}
        \end{center}
      \end{minipage}
      
    \end{tabular}
  \end{center}
  \caption{Strong-scaling calculation time plotted as a function of
    $n_{\rm p}$. Left and right panels show the results in the cases
    of the total number of particles $N$ of $2^{30}$ and $2^{40}$,
    respectively. }
  \label{fig:total_parallel_break_down}
\end{figure}

Figure~\ref{fig:total_parallel_break_down} shows the strong scaling
performance, in the case of the total number of particles ${\rm
  N}$=$2^{30}$ (left panel) and ${\rm N}$ =$2^{40}$ (right panel).  In
the case of $N$=$2^{40}$, $T_{\rm step, para}$ scales well up to
$n_{\rm p}$=$10^6$. On the other hand, in the case of $N$=$2^{30}$,
for $n_{\rm P} \gtrsim 3000$, the slope of $T_{\rm step, para}$
becomes shallower because of $T_{\rm LET,\, p2p}$ becomes relatively
large. For $n_{\rm P} \gtrsim 50000$, $T_{\rm step, para}$ increases
because $T_{\rm LET, allgather}$ becomes relatively large. We can also
see that $T_{\rm step, single}$ increases linearly with $n_{\rm p}$
for $n_{\rm p} \gtrsim 10^5$. This is because $T_{\rm step, single}$
depends on $n_{\rm LET}$ which is proportional to $n_{\rm p}$ for
large $n_{\rm p}$. Thus to improve the scalability further, we need to
reduce $T_{\rm LET, allgather}$ and $T_{\rm LET, p2p}$. We will
discuss the ideas to reduce them in sections~\ref{sec:summary:comm}
and \ref{sec:summary:tree_of_domain}.

\section{Discussion and Summary}

\subsection{Further Reduction of Communication}
\label{sec:summary:comm}

For the simulations on multiple nodes, the communication of the LETs
with neighboring nodes ($T_{\rm LET, p2p}$) would become bottlenecks
for very large $n_{\rm p}$. Thus, it is important to reduce the amount
of the data size for this communication.

An obvious way to reduce the amount of data transfer is to apply some
data compression technique. For example, physical coordinates of
neighboring particles are similar, and thus there is clear room for
compression. However, for many particle-based simulations, compression
in the time domain might be more effective. In the time domain, we can
make use of the fact that the trajectories of most of particles are
smooth. For smooth trajectories, we can construct fitting polynomials
from several previous timesteps. When we send the data of one particle
at new timestep, we send only the difference between the prediction
from the fitting polynomial and the actual value. Since this
difference is many orders of magnitude smaller than the absolute value
of the coordinate itself, we should be able to use much shorter word
format for the same accuracy. We probably need some clever way to use
variable-length word format. We can apply the compression in the time
domain not only for coordinates but for any physical quantities of
particles, including the calculated interactions such as gravitational
potential and acceleration. We can also apply the same compression
technique to communication between the CPU and the accelerator.

\subsection{Tree of Domains}
\label{sec:summary:tree_of_domain}

As we have seen in figure~\ref{fig:total_parallel_break_down}, for
large $n_p$, the total elapsed time increases linearly with $n_{p}$
because the elapsed times for the exchange of LETs and the
construction of the global tree are proportional to $n_{p}$ if $n_p$
is very large. To remove this linear dependency on $n_{p}$, we can
introduce the tree structure to the computational domains (tree of
domains) \citep{Iwasawaetal2018}. By using the tree of domains and
exchanging the locally combined multipole moments between distant
nodes, we can remove {\tt MPI\_Allgather} among all processes to
exchange LETs. It means that the times for the exchange of LETs and
the construction of the global tree do not increase linearly with
$n_p$.

\subsection{Further Improvement on Single Node Performance}

Considering the trends in HPC systems, the overall performance of the
accelerator ($F_{\rm GPU}$ and $B_{\rm GPU}$) increases faster than
the bandwidths of the host main memory ($B_{\rm host}$) and the data
bus between the host and the accelerator ($B_{\rm
  transfer}$). Therefore, in the near future, the main bottleneck of
the performance could be $B_{\rm host}$ and $B_{\rm transfer}$.

The amount of data copy in the host main memory and data transfer
between the host and the accelerator for the reusing step are
summarized in tables~\ref{tab:copy_amount} and
\ref{tab:transfer_amount}. We can see that the amounts of copying data
and transferring data are about $15n$ and $3n$. One reason of these
large amount of data access is that there are four different data
types of particles (FP, EPI, EPJ and FORCE) and data are copied
between different data types. If we use only one data type of
particle, we could avoid to copy data of the procedures e) and g) in
table~\ref{tab:copy_amount} and the procedure B) in
table~\ref{tab:transfer_amount}. If we do so, the amount of copying
data in the main memory and transferring data between the host and the
accelerator could be reduced by 40\% and 33\%, respectively.

\begin{table}
  \begin{tabular}{ll}
    \toprule
     Procedure  & Amount of copying data \\
     \midrule
     a) Calculate multipole moments of local tree  & $n$ \\
     b) Reorder particles for global tree          & $3n$ \\
     c) Calculate multipole moments of global tree & $n$ \\
     d) Copy EPJ and SPJ to the send buffer        & $2n$ \\
     e) Copy EPI to the send buffer                & $2n$ \\
     f) Copy FORCE from the receive buffer         & $2n$ \\
     g) Write back FORCEs to FPs, integrate orbits of FPs and copy FPs to EPIs and EPJs & $4n$ \\
  \bottomrule
  \end{tabular}
  \caption{Amount of coping particle particle data in the host main memory for specified procedure (left column).}
  \label{tab:copy_amount}
\end{table}

\begin{table}
  \begin{tabular}{ll}
    \toprule
     Procedure  & Amount of transferring data \\
     \midrule
     A) Send EPJ and SPJ  & $n$ \\
     B) Send EPI          & $n$ \\
     C) Receive FORCE     & $n$ \\
  \bottomrule
  \end{tabular}
  \caption{Amount of transferring particle data between the host and the accelerator for specified procedure (left column).}
  \label{tab:transfer_amount}
\end{table}

Another way to improve the performance is to implement all procedures
on the accelerator because the bandwidth of the device memory ($B_{\rm
  GPU}$) is much faster than $B_{\rm host}$ and $B_{\rm transfer}$. In
this case, the performance would be determined by the overall
performance of the accelerator.

\subsection{Summary}

In this paper, we described the algorithms we implemented to FDPS to
allow efficient and easy use of accelerators. Our algorithm is based
on Barnes' vectorization, which has been used both on general-purpose
computers (and thus previous versions of FDPS), and also on GRAPE
special-purpose processors and GPGPUs. However, we have minimized the
amount of the communication between the CPU and the accelerator by
indirect addressing method, and we further reduce the amount of
calculation on the CPU side by interaction list reusing. The
performance improvement over the simple method based on Barnes'
vectorization on CPU can be as large as a factor of 10 on the current
generation of accelerator hardware. We can expect fairly large
performance improvement also on future hardware, even if the relative
communication performanceis expected to degrade.

The version of FDPS described in this paper is available at {\tt
  https://github.com/FDPS/FDPS}.

\bigskip

Numerical computations were in part carried out on K computer at RIKEN
Center for Computational Science through the HPCI System Research
project (Project ID:ra000008) and on Cray XC50 at Center for
Computational Astrophysics, National Astronomical Observatory of
Japan. Part of the research covered in this paper was funded by MEXTs
program for the Development and Improvement for the Next Generation
Ultra High-Speed Computer System, under its Subsidies for Operating
the Specific Advanced Large Research Facilities. M.I is supported by
JSPS KAKENHI Grant Number JP18K11334 and JP18H04596.

\appendix
\renewcommand{\thesection}{\Alph{section}}
\renewcommand{\thesubsection}{\Alph{section}.\arabic{subsection}}
\setcounter{section}{0} 
\renewcommand{\theequation}{\Alph{section}.\arabic{equation} }
\setcounter{equation}{0}
\renewcommand{\thetable}{\Alph{section}.\arabic{table}}
\setcounter{table}{0}
\renewcommand{\thefigure}{\Alph{section}.\arabic{figure}}
\setcounter{figure}{0}

\section{Performance model of force kernel on GPGPU}
\label{sec:model_gpu}

\begin{figure}
    \begin{center}
      \includegraphics[width=10cm]{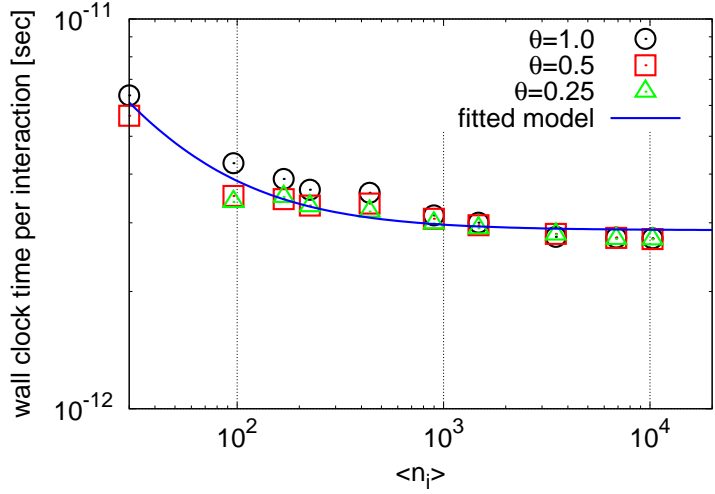}
    \end{center}
    \caption{Elapsed time for interaction calculations per
      interaction.  Circles, squares and triangles indicate the
      results for $\theta$=1.0, 0.5 and 0.25, respectively. Solid
      curve indicates the fitted model.}
  \label{fig:wtime_test_force}
\end{figure}

Here, we construct the performance model of the force kernel on the
GPGPU.  Figure~\ref{fig:wtime_test_force} shows the measured time for
the force kernel per interaction against $\langle n_{\rm i} \rangle$
for various $\theta$. We can see that the elapsed times are
independent of $\theta$ (i.e. independent of $\langle n_{\rm list}
\rangle$) and depend on $\langle n_{\rm i} \rangle$. For small
$\langle n_{\rm i} \rangle$, the time decreases as $\langle n_{\rm i}
\rangle$ increases. This is because the times are determined by the
bandwidth of the main memory of GPGPUs ($B_{\rm GPU}$). For large
$\langle n_{\rm i} \rangle$, the elapsed times are almost constant
because these times are determined by the speed of the floating point
operation of GPGPUs ($F_{\rm GPU}$). Thus the time for the force
kernel $T_{\rm kernel}$ is given by
\begin{eqnarray}
  T_{\rm kernel} &=& n \langle n_{\rm list} \rangle \left( \frac{\alpha_{\rm GPU\, calc} n_{\rm op}}{F_{\rm GPU}}
  + \frac{\alpha_{\rm GPU\, transfer} \left(b_{\rm EPJ}+b_{\rm ID} \right)}{ \langle n_i \rangle {B_{\rm GPU}}}  \right).
  \label{eq:model_kernel}
\end{eqnarray}

To determine the coefficients for equation~(\ref{eq:model_kernel}), we
assume that $F_{\rm GPU}$ is 13.8 Tflops and ${B_{\rm GPU}}$ is about
550 GB/s, which is measured with {\tt bandwidthTest} in the NVIDIA
SDK. These coefficients are listed in table~\ref{tab:model_kernel}.

\begin{table}
  \caption{Coefficient for equation~(\ref{eq:model_kernel})}
  \label{tab:model_kernel}
  \begin{tabular}{ll}
    \toprule
    $\alpha_{\rm GPU\, calc}$ & 1.7 \\
    $\alpha_{\rm GPU\, transfer}$ & 2.7   \\
    $n_{\rm op}$ & 23  \\
  \bottomrule
  \end{tabular}
\end{table}

\section{Performance of {\tt MPI\_Gather} and {\tt MPI\_Allgather} on K Computer}
\label{sec:allgather}

Here, we construct the performance models of {\tt MPI\_Gather} and
{\tt MPI\_Allgather} on K computer. On K computer, the performance of
{\tt MPI\_Gather} is almost the same as that of {\tt MPI\_Allgather}.
Thus in the following, we only consider {\tt MPI\_Allgather}.

The elapsed time for {\tt MPI\_Allgather} $T_{\rm allgather}$ as a
function of message size $b$ and the number of processes $n_{\rm p}$
is given by
\begin{equation}
  T_{\rm allgather} (b, n_{\rm p}) = T_{\rm allgather, startup} (n_{\rm p}) + T_{\rm allgather, words} (b, n_{\rm p}),
\end{equation}
where $T_{\rm allgather, startup}$ is the start up time which depends
on only $n_{\rm p}$ and $T_{\rm allgather, words}$ is the time for
transfer of the message which depend on both $n_{\rm p}$ and $b$. For
small message size, $T_{\rm allgather} \sim T_{\rm allgather,
  startup}$. Thus to determine $T_{\rm allgather, startup}$ we
measured the times for {\tt MPI\_Allgather} with a short message size.

\begin{figure}
    \begin{center}
      \includegraphics[width=10cm]{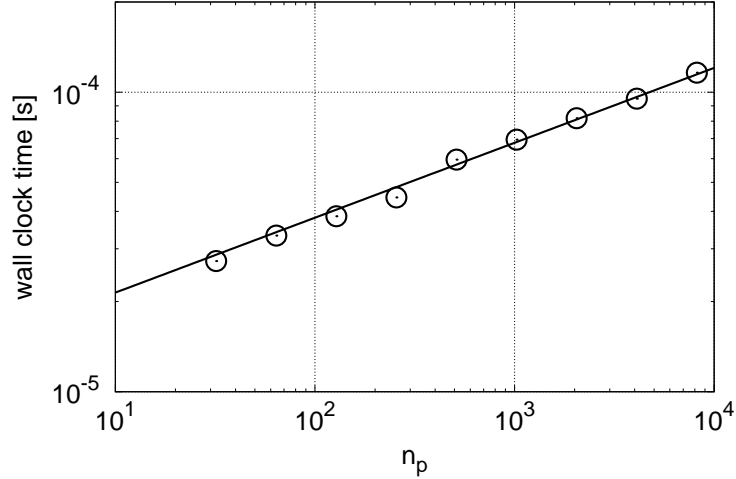}
    \end{center}
  \caption{Elapsed time for {\tt MPI\_Allgather} to send message of 2
    bytes against the number of processes. Circles indicate the
    measurement values and solid curve indicates our fitting model.}
  \label{fig:wtime_allgather_startup}
\end{figure}

Figure~\ref{fig:wtime_allgather_startup} shows the elapsed time for
{\tt MPI\_Allgather} to send message of two bytes against $n_{\rm
  p}$. We can see that $T_{\rm allgather, startup}$ is proportional to
$n_{\rm p}^{1/4}$. Thus $T_{\rm allgather, startup}$ is given by
\begin{equation}
  T_{\rm allgather, startup} \sim n^{1/4}_{\rm p} \tau_{\rm allgather, startup}.
\end{equation}

For large message size, $T_{\rm allgather}$ should be determined by
the injection bandwidth $B_{\rm inj}$. Thus $T_{\rm allgather, words}$
is give by
\begin{equation}
  T_{\rm allgather, words} \sim \frac{\alpha_{\rm allgather} b n_{\rm p}}{B_{\rm inj}}.
\end{equation}

Thus the elapsed time for {\tt MPI\_Allgather} is given by
\begin{equation}
  T_{\rm allgather} \sim n^{1/4}_{\rm p} \tau_{\rm allgather, startup} + \frac{\alpha_{\rm allgather} b n_{\rm p}}{B_{\rm inj}}.
  \label{eq:t_allgather}
\end{equation}

\begin{figure}
    \begin{center}
      \includegraphics[width=10cm]{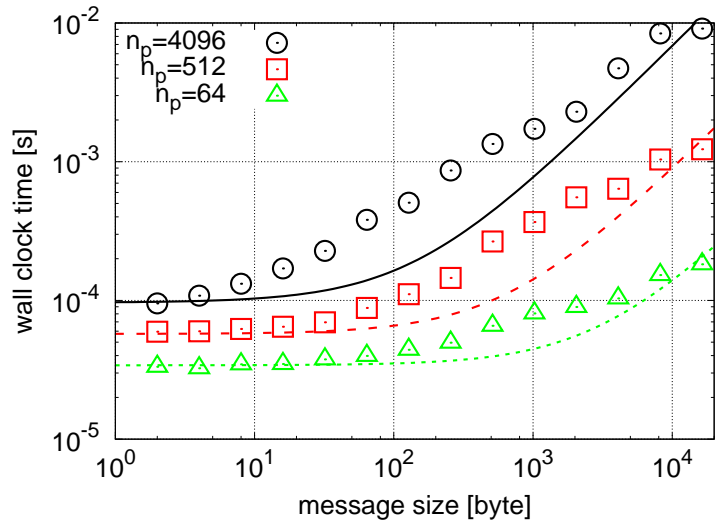}
    \end{center}
  \caption{Elapsed times for {\tt MPI\_Allgather} plotted against the
    message size. Circle, square and triangle indicate the result with
    $n_{\rm p}$=4096, 512 and 64, respectively. Solid, dashed and
    dotted curves indicate our fitting models for $n_{\rm p}$=4096,
    512 and 64, respectively.}
  \label{fig:wtime_allgather}
\end{figure}

Figure~\ref{fig:wtime_allgather} shows the measured and predicted
times for {\tt MPI\_Allgather} against the message size. Here, we
assume $B_{\rm inj}$=4.8 GB/s from the result of point-to-point
communication test. The parameters in equation~(\ref{eq:t_allgather})
are listed in table~\ref{tab:t_allgather}. Our model agrees with the
measured data.

\begin{table}
  \caption{Coefficient for equation~(\ref{eq:t_allgather})}
  \label{tab:t_allgather}
  \begin{tabular}{ll}
    \toprule
    $\tau_{\rm allgather, startup}$  & $ 1.2 \times 10^{-5}$ sec  \\
    $\alpha_{\rm allgather}$       & 0.62 \\
  \bottomrule
  \end{tabular}
\end{table}

\end{document}